\documentclass[a4paper,11pt]{article}

\usepackage[a4paper,
  left=2.5cm, right=2.5cm,
  top= 3cm, bottom=4cm]{geometry}
\usepackage{amsmath}
\usepackage{oldgerm}
\usepackage{amssymb}
\usepackage{bbm}
\usepackage[dvips]{graphicx}
\usepackage{epsfig}
\usepackage{color}
\usepackage{cite}
\usepackage{epic}
\usepackage{hyperref}
\usepackage{multirow}

\usepackage[T1]{fontenc}
\usepackage[english]{babel}
\numberwithin{equation}{section}

\newcommand{\be}{\begin{equation}}  
\newcommand{\ee}{\end{equation}}

\newcommand{\rem}[1]{} 

\def\Hirz[#1]{\mathbbm{F}_{#1}}
\def\o[#1]{\overline{#1}}

\def\ad3{\ensuremath{\overline{\rm{D3}}}}
\def\vo{\mathcal{V}}

\newcommand{\todo}[1]{}
\renewcommand{\todo}[1]{{\color{red} Comment: {#1}}}

\newcommand{\vol}{\mathcal{V}}

\newcommand{\K}{K\"{a}hler }

\begin{document}

\begin{titlepage}

\vskip -0.5cm

\begin{flushright}

\end{flushright}
 
\vskip 1cm
\begin{center}
 
{\Large \bf On de Sitter String  Vacua from Anti-D3-Branes in the  \\ Large Volume Scenario} 
 
 \vskip 1.2cm
 
{\small{\bf Chiara~Crin\`o,$^{a,b}$  Fernando Quevedo$^c$  and Roberto~Valandro$^{a,b}$}}

 \vskip 0.4cm
 
{\small{\it $^{a}$ Dipartimento di Fisica, Universit\`a di Trieste,  Strada Costiera 11,\\ I-34151 Trieste, Italy \\ }}
\vspace{.1cm}
{\small{\it $^{b}$  INFN, Sezione di Trieste, Via Valerio 2, I-34127 Trieste, Italy \\ }}
 \vspace{.1cm}
 {\small{\it $^{c}$ DAMTP, University of Cambridge, Wilberforce Road, \\ Cambridge, CB3 0WA, UK
 }}
 \vskip 0.5cm
 {\tiny   \texttt{chiara dot crino at ts dot infn dot it, f dot quevedo at damtp dot cam dot ac dot uk, \\ \vspace{-2mm} roberto dot valandro at ts dot infn dot it}}
 \vskip 1.5cm

\abstract{\small{We consider de Sitter vacua realised in
concrete type IIB Calabi-Yau compactifications with an anti D3-brane at the tip of a warped throat of  Klebanov-Strassler type. 
The K\"ahler moduli are stabilised together with the complex structure modulus of the warped throat. The volume is exponentially large as in the large volume scenario (LVS). We analyse the conditions on the parameters of the EFT such that they are in the region of validity of our approximations, there are no runaway problems and the vacua satisfy all consistency constraints, such as tadpole cancellation.
We illustrate our results with an explicit Calabi-Yau orientifold with two  K\"ahler moduli and one antibrane on top of an O3-plane in a warped throat, that has the goldstino as its only massless state. The moduli are stabilised with $g_s\sim 0.2$ and volume $\mathcal{V}\sim 10^4$ in string units, justifying the approximation used to derive the corresponding EFT. Although the model lacks chiral matter, it is presented as a proof of concept, chosen to be the simplest realisation of antibrane uplift.
} }

\end{center}

\end{titlepage}

\tableofcontents

\newpage

\section{Introduction}

It has been known for more than 20 years that our universe is in a period of accelerated expansion. This observation has led to many efforts to obtain de Sitter (dS) space from string theory. The first successful attempt was the KKLT scenario \cite{Kachru:2003aw} \, in which the introduction of antibranes in type IIB flux compactifications is a crucial ingredient to stabilise  all moduli  in a metastable de Sitter vacuum. These flux compactifications lead to the string theory realisation  of a landscape of  vacua that is at the moment the main approach towards explaining the smallness of the cosmological constant \cite{Bousso:2000xa}. 

Given the magnitude of the problem of fixing all moduli at a positive vacuum energy,  the  scenario naturally requires several non-trivial ingredients of string compactifications: fluxes, orientifolds, non-perturbative effects, antibranes, etc. The combination of all these ingredients in a consistent way should be subject to close scrutiny since the physical implications are so important. This is precisely what has been happening during the past 15 years. Several criticisms have been pointed out to this approach, ranging from the explicitness of the scenario to the proper use of effective field theory (EFT), consistency of the antibrane together with non-perturbative effects, etc. This has led to perspectives as different as the claims that there is an almost infinite discrete number of solutions to the ones that there are no solutions at all. So far, antibrane uplift has survived all challenges which is helping to the robustness of the scenario. Our goal in this paper is to add to this robustness by constructing a concrete model exhibiting all the ingredients of the scenario and, expanding on \cite{Garcia-Etxebarria:2015lif}, explicitly stabilising the moduli with positive vacuum energy.\footnote{Previous concrete examples of de Sitter uplift in type IIB orientifold compactification have been obtained in which the uplift is not due to the presence of antibranes but to other sources, such as T-branes, $\alpha'$-corrections or  D-term generated racetrack potrential \cite{Cicoli:2012vw,Cicoli:2013mpa,Cicoli:2013cha,Cicoli:2015ylx,Cicoli:2017shd, Gallego:2017dvd,Louis:2012nb,Braun:2015pza}.}


In the KKLT scenario the anti D3-brane is typically placed at the tip of a highly warped throat, so that its contribution to the vacuum energy  is just a perturbation of the supergravity potential of $\mathcal{N}=1$ 4d compactification (not spoiling the moduli stabilisation present already without the anti D3-brane). 
In particular we are interested in an anti D3-brane on top of an O3-plane at the tip of a warped throat with (2,1) three-form fluxes. In this case, in fact, the goldstino is the only low-energy degree of freedom. This justifies the use of a nilpotent superfield $X$ to describe the EFT \cite{Kallosh:2015nia}. 

In \cite{Garcia-Etxebarria:2015lif} this problem was addressed in a geometry obtained by orientifolded conifolds, refining and generalising the analysis in \cite{Kallosh:2015nia}.
Already the standard conifold singularity can support an orientifold involution necessary to produce an O3-plane at the tip of the throat. Deforming the conifold singularity leads to two O3-planes placed at the north and south poles of the blown up  $S^3$. 
The interesting limit is when the two O3-planes come close (but not on top of each other); this is controlled by the complex structure modulus giving the size of the $S^3$. 
If in this region an imaginary self dual 3-form flux is present, a warped throat is generated \cite{Klebanov:2000hb,Giddings:2001yu} and the O3-planes sit at the bottom of the throat. 
In \cite{Garcia-Etxebarria:2015lif}, a procedure to pick up CY threefolds was outlined and used to find explicit models with the desired throat and involution. 
We apply the method of \cite{Garcia-Etxebarria:2015lif} to construct an explicit model with two O3-planes at the tip of a warped throat. 
Including the contribution of the anti D3-brane to the scalar potential, we stabilise the moduli, finding a de Sitter minimum with large volume as in LVS  \cite{Balasubramanian:2005zx,Conlon:2005ki}.

Before constructing the model, we study moduli stabilisation for type IIB orientifold compactifications on CY's with two K\"ahler moduli and a Klebanov-Strassler (KS) \cite{Klebanov:2000hb} throat. Following \cite{Bena:2018fqc,Blumenhagen:2019qcg} we consider a three-moduli potential, where the third scalar is the complex structure modulus controlling the size of the $S^3$ and the warp factor at the tip of the throat (the remaining complex structure moduli are assumed to be stabilised by fluxes at higher scales). This inclusion turns out to be unnecessary: our results do not deviate much from what found in \cite{Aparicio:2015psl}, where the authors assumed that the warped factor was fixed at the KS value, like in \cite{Giddings:2001yu}.

From our analysis, it turns out that stabilising the moduli in such setups is not easy at all, as several constraints must be satisfied. In this paper we present a generic analysis that shows a tension between the several requirements. These come from asking control over approximations and the presence of a de Sitter minimum realized by an \ad3 at the tip of a warped throat. The example we discuss at the end of the paper fulfills all the consistency conditions.

One example of such a condition was recently discovered in  \cite{Bena:2018fqc,Blumenhagen:2019qcg,Dudas:2019pls}:  a runaway may appear if the flux number $M$ (that is $F_3$ 3-form flux along the $S^3$ at the tip of the throat) is not sufficiently large; they found the bound $g_sM^2\gtrsim 47$. 
Actually, we stress that the requirement of control over the Klebanov-Strassler approximation of the warped throat, i.e. $g_s|M|\gg 1$,  imposes a much  larger lower bound on $M$ in the perturbative regime.\footnote{One may improve the bound on $M$ by making $g_s$ larger (still remaining in the perturbative regime). However in LVS \cite{Balasubramanian:2005zx,Conlon:2005ki}, at the minimum $g_s$ is related to the volume of the compactification manifold, that for larger $g_s$  tends to be not so large, risking a violation also of the sugra approximation.} 

This bound on $M$ translates to a condition on the D3-tadpole. In fact, the fluxes $M$ and $K$ along the throat contribute by $MK$ to the total positive D3-charge. Since the warp factor ($\sim e^{-8\pi K/3g_sM}$) should be small, a lower bound on $M$ puts also a lower bound on $K$ and then on $MK$. Having a large positive D3-charge is potentially problematic, since the negative contribution is fixed by the geometry of the D7-brane and O7-planes in the setup (and possible O3-planes). Moreover further (positively contributing) fluxes should be added to fix the complex structure moduli {\it \`a la} GKP \cite{Giddings:2001yu}. This means that a large lower bound on $MK$ is a serious issue in type IIB compactifications.\footnote{ Recently a further tension between D3-tadpole cancellation and complex structure moduli stabilisation has been discussed in \cite{Bena:2020xrh} (and observed in a specific example in \cite{Braun:2020jrx}).} However it is also important to stress that large negative D3-charge objects are available also in perturbative regime (like Whitney branes \cite{Collinucci:2008pf}). We will in fact show in the explicit example that such objects allow to satisfy D3-tadpole cancellation condition also for large $MK$ satisfying the bound.

The paper is organised as follows. In Section~\ref{sec:typeIIBmdstab} we present the background material for flux compactifications, LVS and KKLT warped antibrane uplift, including the role of nilpotent superfields; we also review the different challenges that have been raised over time for antibrane uplift. In Section~\ref{sec:2steps} we analyse in detail the stabilisation of the K\"ahler moduli and the complex structure modulus of the warped throat: we write the explicit three-moduli scalar potential and we derive the minimum conditions. In Section~\ref{sec:boundsond3tadpole} we discuss the consistency conditions the minima have to fulfill to correspond to a consistent string model and also for the validity of the corresponding EFT; we also explain their impact on the D3-tadpole. In Section~\ref{Sec:ExplicitModel} we show a concrete CY compactification that produces a de Sitter minimum of the type considered in Section~\ref{sec:2steps} and that fulfills all the requirements listed in Section~\ref{sec:boundsond3tadpole}. We write our conclusions in Section~\ref{sec:Concl}.

\section{Type IIB moduli stabilisation with anti D3-branes}\label{sec:typeIIBmdstab}

Compactifying type IIB string theory on a Calabi-Yau (CY) $X$ in presence of O-planes, one obtains a low energy  effective field theory described by an $\mathcal{N}=1$ four dimensional (4d) supergravity (sugra). 

If no further ingredients are added to the compactification, there are several massless scalars, called {\it moduli}, coming from reducing the 10d type IIB fields $g_{10}$ (metric), $C_0,C_2,C_4$ (RR p-form potentials), the NSNS two-form $B_2$  and the dilaton $\phi$. The orientifold projection separates each cohomology group $H^{p,q}(X)$ into an even and an odd part with dimensions respectively $h^{p,q}_+$ and $h^{p,q}_-$. There are also moduli coming from the open string spectrum, that will not be analysed in this paper. The closed string moduli are:
\begin{itemize}
	\item  The axio-dilaton: $S=e^{-\phi}-iC_0$. 
	\item $h^{1,1}_-$ complex scalars coming from $C_2-iS\,B_2$.
	\item $h^{1,1}_+$ complex scalars $T_i(x)$, called \K moduli, corresponding to deformations of the \K form $J$ ($\tau_i$) and of the RR $C_4$ ($\theta_i$).\footnote{They also depend on the periods of $B_2$ and $C_2$ on odd two-cycles. However, in this paper we will consider models with $h^{1,1}_-=0$.} 
	\item $h^{1,2}_-$ complex scalars  $Z_\alpha(x)$, called complex structure moduli,  corresponding to deformations of the complex structure of $X$. 
\end{itemize}
There are also  $h^{1,2}_+$ 4d massless vectors coming from the reduction of $C_4$ over three-forms.
In this paper we will consider models with $h^{1,1}_-=0$, hence the scalars from $B_2,C_2$ will not play a role.

The $\mathcal{N}=1$ 4d effective theory is described by the \K potential
\begin{equation}
	K_0(T_i, Z_\alpha, S)=-2 \ln (\vol)-\ln (S+\bar{S})-\ln \left(-i \int \Omega \wedge \bar{\Omega}\right) 
\end{equation}
where the volume $\vol$ of the CY  depends on the K\"ahler moduli $T_i$ and the 
holomorphic (3,0)-form $\Omega$ depends only on the complex structure moduli $Z_\alpha$.

\subsection{Fluxes, complex structure stabilisation  and D3-tadpole}

The axio-dilaton $S$ and all the complex structure moduli $Z_\alpha$ can be stabilised at the classical level, by turning on a non-zero vev for the field strength $G_3=F_3-iSH_3$ of the RR and NSNS 2-form potential  $C_2, B_2$. The three-forms $F_3,H_3$ have quantised fluxes over the integral three-cycles $\Sigma_A$ ($A=1,...,b_3$) of the CY threefold $X$: 
\begin{equation}
	\frac{1}{(2\pi)^2\alpha'}\int_{\Sigma_A}F_3=M_A; \quad \frac{1}{(2\pi)^2\alpha'}\int_{\Sigma_A}H_3=-K_A \quad\mbox{with }\,\, M_A,K_A\in \mathbb{Z}\:.
\end{equation}
These fluxes induce a D3-charge
\begin{equation}
	Q_{D3}^{\rm flux} = \frac{1}{(2\pi)^4\alpha'^2}\int F_3\wedge H_3\:.
\end{equation}
The equation of motion implies the 3-form $G_3$ to be imaginary self dual\cite{Dasgupta:1999ss,Giddings:2001yu}; this implies the corresponding $Q_{D3}^{\rm flux}$ to be positive. The total D3-charge must be zero in order to cancel  the D3-tadpole. In a CY orientifold compactification of type IIB string theory there are negative sources of D3-charge: O3-planes, O7-planes and D7-branes \cite{Giddings:2001yu}. The contribution of the last two is captured in the F-theory description by the famous $\frac{\chi}{24}$, where $\chi$ is the Euler characteristic of the fourfold. This contribution can actually be large, also in perturbative type IIB compactifications, as we will see in the explicit model in Section~\ref{Sec:ExplicitModel}.

The flux $G_3$ induces a non-zero superpotential in the 4d $\mathcal{N}=1$ effective theory\cite{Gukov:1999ya}:
\begin{equation}
	W_{flux}(Z_\alpha, S)=\int G_3\wedge\Omega  \:.
\end{equation} 
The corresponding F-term potential is,
\begin{equation}
	V=e^{K_0}\left[K_0^{I \bar{J}} D_{I} W_{flux} \overline{D_{J} W_{flux}}\right]
\end{equation}
where  $I,J=Z_\alpha, S$ and $D_{I} W_{0} \equiv \partial_{I} W_{0}+\left(\partial_{I} K\right) W_{0}$.
This makes the potential be positive semi-definite; hence it has a minimum at $V=0$, corresponding to $D_{S,Z_\alpha}W_0=0$. The contribution of the K\"ahler moduli vanishes due to the no-scale structure. At the minimum of $V$ the K\"ahler moduli are flat directions.


\subsection{LVS K\"ahler moduli stabilisation}\label{Section:LVSmoduliStab}

In order to stabilise the \K moduli, one needs to include quantum effects. There are mainly two scenarios in which this has been achieved: KKLT \cite{Kachru:2003aw} and LVS \cite{Balasubramanian:2005zx, Conlon:2005ki}. There are also mixed situations \cite{Westphal:2006tn,Rummel:2011cd,AbdusSalam:2020ywo}. In this paper we will concentrate on the Large Volume Scenario (LVS). Here, the $\alpha'$ corrections to the K\"ahler potential compete with non-perturbative corrections to the  superpotential to stabilise the K\"ahler moduli. 

After including the $
\alpha'$ corrections, the \K potential for the K\"ahler moduli  reads:
\begin{equation}\label{eq:LVSKpot}
	K=-2 \ln \left[\mathcal{V}+\frac{\xi}{g_{s}^{3/2}}\right]\:,
\end{equation}
where $ \xi \equiv-\frac{\chi(X) \zeta(3)}{4(2 \pi)^{3}}$. $\chi(X)$ is the Euler characteristic of the CY threefold, $\zeta(3)\sim 1.202$. 
In \cite{Minasian:2015bxa}, contributions coming from the O7/D7 sector have been claimed to modify the constant $\xi$ by replacing $\chi(X)\mapsto \chi(X) + 2\int_X D_{O7}^3$.

The superpotential is
\be
W=W_0 + \sum_i A_i e^{-a_i T_i}\:,
\ee
where $W_0$ is a constant term, that is the vev of the tree-level superpotential $W_{flux}$ generated by the background fluxes $F_3,H_3$, and the second term is the non-perturbative contribution.
The real part of the modulus $T_i$ is the volume of an internal four-cycle wrapped either by an euclidean D3-brane (E3)  (in which case $a_i=2\pi$) or by a stack of D7-branes supporting a condensing gauge theory ($a_i= 6\pi/b_0$ with $b_0$ the coefficient of the one- loop beta function). 
 In the following we consider only two values for $a$: $a=2\pi$ when the superpotential is generated by an E3 instanton;\footnote{One may also consider situations where the leading contribution is given by an E3 that is effectively wrapping $n$-times the non-perturbative divisor (see \cite{Berglund:2012gr}). In this case $a=2\pi n$.} $a=\frac{\pi}{3}$ when it is generated by gaugino condensation for an $SO(8)$ gauge group.  In perturbative string theory, $SO(8)$ is the only condensing gauge group on a D7-brane stack with no (chiral and non-chiral) matter; this happens if it wraps a rigid cycle with $h^{1,0}=0$. All $SU(N)$ gauge group, that would have $a=\frac{2\pi}{N}$, must have non-zero intersections with other D7-branes, having at best vector-like matter that may spoil gaugino condensation.\footnote{We are grateful to Luca Martucci for a detailed discussion on this point.}

In this paper we consider  situations in which only a single non-perturbative effect is allowed, giving a superpotential
\be
W=W_0 +A e^{-a T_s}\:.
\ee
Notice that the perturbative correction violate the no-scale condition.

Let us consider a CY threefold $X$ with $h^{1,1}=2$ and a Swiss cheese form of the volume, i.e.
\begin{equation}\label{SCvolume}
	\vol=\kappa_b \tau_b^{3/2}-\kappa_s\tau_s^{3/2}
\end{equation}
where $\tau_b$ and $\tau_s$ are the volumes of two 4-cycles of $X$; $\kappa_b, \kappa_s$ are constants depending on the detailed intersection form on $X$. We will set $\kappa_b$ equal to 1 in the following (by an appropriate rescaling of the modulus $\tau_b$).
The LVS scalar potential is then\footnote{By generality argument, we have taken $e^{K_{c.s.}}\simeq 1$.}
\begin{equation}
	V_{LVS}=\frac{8 g_s A^{2} a^{2} \sqrt{\tau_{s}} e^{-2 a \tau_{s}}}{3\kappa_s \vol}+\cos \left(a \theta_{s}\right) \frac{4 g_s A a W_{0} \tau_{s} e^{-a \tau_{s}}}{\vol^2}+\frac{3 W_{0}^{2} \xi}{2g_s^{1/2}\vol^3} \:.
\end{equation}
It has a non-supersymmetric minimum at
\begin{subequations}
	\begin{align}
	\theta_s &= \frac{\pi}{a}\:,\\
	\tau_b^{3/2} &=\frac{3 W_0\kappa_s\sqrt{\tau_s}e^{a\tau_s}}{a A}\frac{(1- a \tau_s)}{(1-4 a\tau_s)} \simeq \frac{3 W_0\kappa_s\sqrt{\tau_s}e^{a\tau_s}}{4a A} \:,\\
	\tau_s^{3/2} &=\frac{\xi}{16 a \tau_s \kappa_s g_s^{3/2}}\frac{(1-4 a \tau_s)^2}{(a \tau_s-1)} \simeq \frac{\xi}{ g_s^{3/2}\kappa_s}\:,
	\end{align}
\end{subequations}
where the approximated results on the r.h.s. are obtained under the assumption $a\tau_s\gg 1$, that must be assumed to ignore higher instanton corrections to the non-perturbative superpotential, as we have done (the sugra approximation requires moreover $\tau_s$ to be much bigger than 1).
We see that $\tau_b\gg\tau_s$ and that the volume $\vol\simeq \tau_b^{3/2}$ is exponentially large.
The value of the potential at the minimum is 
\begin{equation}
	V_{\min }=-\frac{12 g_s W_{0}^{2} \kappa_s \tau_{s}^{3 / 2} \left(a \tau_{s}-1\right)}{\tau_{b}^{9 / 2}\left(4 a \tau_{s}-1\right)^{2}}  
	\simeq -\frac{3 g_s W_{0}^{2} \kappa_s\tau_{s}^{1 / 2}}{4a\tau_{b}^{9 / 2}}    <0 \:.
\end{equation}
Hence, the minimum is Anti-de Sitter (AdS).

\subsection{Anti D3-branes, Uplift Term and Nilpotent Goldstino}

The three-form fluxes, that stabilise the complex structure moduli, back-react to the geometry by  inducing a warp factor $e^{2D}$ in the 4d metric as well as a conformal factor in front of the CY internal space:
\begin{equation}
ds^2=e^{2D} ds_4^2+ e^{-2D} ds_{CY}^2 \:,
\end{equation}
where $D$ depends on the internal coordinates.
The regions of the internal metric where $e^{-2D}$ is very large are called {\it warped throats}. Correspondingly the warp factor is very small and the 4d scales are strongly redshifted.
One can write
$e^{-4D}=1+\frac{e^{-4A}}{\vo^{2/3}}$. A large warped throat 
is then a region where $e^{-2A}\gg \vo^{1/3}$. 
The warp factor depends on the complex structure of the CY threefold.

These throats typically arise  in correspondence with deformed conifold singularities~\cite{Giddings:2001yu}. The blown up $S^3$ that deforms the conifold singularity sits at the tip of the throat. Its volume is controlled by the complex structure modulus $Z$, that is the period of the holomorphic (3,0)-form over the $S^3$. 
The factor $e^{4A}$ at the tip of a long warped throat depends  on the stabilised value of $Z$. If one switches on a
flux $M$ of $F_3$ over the $S^3$ at the tip and a flux $K$ of $H_3$ over the dual cycle, then the leading contribution to the warp factor is~\cite{Giddings:2001yu}:  
\be \label{eq:warpFactorTip}
e^{4A_0} \simeq e^{-8\pi K/3g_sM} \:.
\ee
Tuning $M,K$ such that $e^{4A_0} \ll 1$, produces at the same time a small warp factor (high redshift) and a long throat. When the supergravity description is a valid approximation, i.e. when the size of the $S^3$ at the tip of the conifold is larger than the string length, then the metric in the throat can be approximated by the Klebanov-Strassler (KS) solution \cite{Klebanov:2000hb}.
Requiring a large volume for the $S^3$ at the tip is equivalent to asking
\begin{equation}\label{sugragsMbound}
 g_s |M| \gg 1 \:.
\end{equation}

In the KS solution, the three-form flux $G_3$ along the throat is imaginary-self dual. It contributes to the D3-tadpole by
\be
Q_{D3}^{\rm (KS flux)} = M\,K
\ee
that is necessarily a positive number for an imaginary self dual flux (without loss of generality, in the following we will always consider positive values for $M$ and $K$). 
Notice that, since the total $G_3=G_3^{\rm throat}+G_3^{\rm rest}$ must be imaginary self dual, then  the flux $G_3^{\rm rest}$ in the bulk must be imaginary self dual as well (and then it also contribute positively to the D3-charge).

Without any further ingredient, the LVS vacuum would be AdS. To have a dS minimum one needs to introduce a source giving a positive contribution to the scalar potential. In this paper we will follow  \cite{Kachru:2003aw} and add an anti D3-brane (\ad3) at the tip of a throat. Its contribution to the vacuum energy will be redshifted by the warp factor.
The contribution to the energy of the \ad3-brane is
\cite{Kachru:2003sx} (see also \cite{Blumenhagen:2019qcg,Dudas:2019pls} for a recent review in the context of dS moduli stabilisation)
\begin{equation}\label{VupAntiD3}
V_{\overline{D3}} =2 T_3 e^{4 D(0)}\simeq \frac{e^{4A_0}}{ \vol^{4/3}} M_p^4 \ll M_s^4\:,
\end{equation}
where   $\mathcal{V}$ is the volume of the extra dimensions and $M_s$ and $M_p$ are the string and Planck scale respectively. 

The imaginary self dual 3-form flux gives mass to some of the  $\ad3$ modes, leaving only a $U(1)$ gauge field and one single fermion (goldstino) in the massless spectrum. By introducing an orientifold projection, one can project out the gauge field and  keep
only the goldstino in the spectrum.
In  \cite{Kallosh:2015nia,Garcia-Etxebarria:2015lif}, it has been found how to choose an orientifold involution, such that there is an O3-plane at the tip of the throat. 
When an \ad3-brane is placed on top of such an O3-plane, one can describe the $\ad3$ degrees of freedom by a nilpotent superfield $X$ (i.e. $X^2=0$).
The nilpontency condition implies a constraint on the components of the chiral superfield $X$, where
\be
X=X_0(y)+\sqrt{2}\psi(y) \theta + F(y)\theta\bar{\theta} \:,
\ee
with, as usual, $y^\mu=x^\mu+ i \theta \sigma^\mu \bar{\theta}$. In fact, imposing $X^2=0$  implies 
$X_0=\frac{\psi\psi}{2F}$.

The representation in terms of a nilpontent superfield $X$  
is very convenient since it allows to treat 
the  effect of an \ad3-brane in terms of standard supergravity couplings of matter and moduli superfields to the nilpotent goldstino (see for instance \cite{Antoniadis:2014oya, Ferrara:2014kva, Kallosh:2014via, Kallosh:2015nia, Bergshoeff:2015jxa,Bandos:2015xnf,Bertolini:2015hua,Aparicio:2015psl,Bandos:2016xyu}). 
Notice that the lowest component of $X$ has zero vev.
In particular, when the \ad3-brane is on top of the O3-plane,  the modulus describing its motion is absent, contrary to D3-branes in the bulk. This fits with the fact that the scalar component of $X$ is not a propagating field.
Moreover, in calculating the scalar potential, there is no contribution from $X_0$: it is consistently set to zero when looking for Lorentz preserving vacuum configurations as we set all fermions to zero. 

The  K\"ahler potential and the superpotential describing the EFT (after integrating the complex structure moduli and the axiodilaton) are then modified by adding the following terms  
\begin{equation}
K_X = \frac{X \bar{X}}{\vo^{2/3}} \qquad\mbox{and}\qquad W_X =\eta \,X \:.
\end{equation}
Here $\eta= \frac{Z^{2/3}iS}{M}\sqrt{\frac{c''}{\pi}}$ \cite{Dudas:2019pls}, with $Z$ the complex structure of the throat, $M$ the 3-form flux, $c''\equiv \frac{2^{1/3}}{\mathcal{I}(0)}\approx 1.75$ \cite{Bena:2018fqc} and $S$ the axio-dilaton. As we will see below, these terms lead to the uplift term \cite{Dudas:2019pls}
\begin{equation}\label{uplifttermDudasEtAl}
 V_{up} =M_p^4 \frac{g_s|\eta|^2}{ \mathcal{V}^{4/3}} = M_p^4 \frac{c''}{\pi\,g_sM^2}\frac{|Z|^{4/3}}{\vol^{4/3}}    \:.
\end{equation}

\subsection{Challenges to the antibrane uplift scenario}

As we mentioned in the introduction there has been several challenges regarding the validity and robustness of the KKLT antibrane uplift scenario. Here we briefly review them.

\begin{itemize}
\item{\it Moduli stabilisation is non generic.}  Since the mid 1980's it is known that the natural vacuum state is the zero string coupling $g_s$ and infinite string volume $\mathcal{V}$ solution. That is 10d string theory.  Any solution in 4d and finite coupling will have to lead asymptotically to infinite volume and zero coupling. We know that $1/\mathcal{V}$ and $g_s$ are the expansion parameters for $\alpha'$ and string loop expansions respectively. This means that if we want any calculation to be under full control, i.e. with arbitrary precision in string and $\alpha'$  expansions, we have to be in a runaway regime where both expansion parameters go to zero. This is the celebrated Dine-Seiberg problem \cite{Dine:1985he}. This has been recently  revived in terms of the swampland conjectures \cite{Palti:2019pca}. The only way out, as Dine and Seiberg emphasised originally, is to have other parameters that may be present in string compactifications, such as the rank of the gauge groups $N$ or other large integers that may exist. Flux compactifications in IIB string theory actually provide these parameters very naturally. In fact, for compactifications with hundreds of complex structure moduli, there are many of these integers that can take very different  values possibly leading to weak coupling solutions. In this sense flux compactifications provide the best way to address this challenge even though full calculation control cannot be achieved within any compactification. 

We may summarise the situation as follows. For tree-level K\"ahler and superpotentials $K_0,W_0$ the scalar potential vanishes. They will both get quantum corrections $\delta K$ and $\delta W$ with $\delta W$ only coming from non-perturbative effects. The scalar potential gets modified as:
\begin{equation}
V=\delta V\sim e^K\left(W_0^2 \delta K +W_0\delta W\right) \:.
\end{equation} 
Generically the first term dominates since $\delta K$ is perturbative whereas $\delta W$ is non-perturbative, giving the standard runaway behaviour of the Dine-Seiberg problem. However, if it is possible to  achieve $W_0\delta K\sim \delta W$ a non-trivial minimum can exist. This happens either if $W_0\ll 1$ as in KKLT \footnote{Recently, explicit mechanisms have been found that can give rise to exponentially small flux superpotentials $W_0$ \cite{Demirtas:2019sip,Blumenhagen:2020ire,Demirtas:2020ffz}.} or if $\delta K\sim \delta W$ as in LVS.
 
Notice  that having runaway to be the most generic solution should {\it not} be confused with quintessence. Quintessence needs a very particular and tuned runaway in which the fastest rolling direction is extremely tuned to be slow enough as not to contradict experimental constraints on varying constants of nature and fifth forces. In this sense getting quintessence is at least as difficult as getting de Sitter. Furthermore, it should be stressed that  not having full computational control does not mean that there is no control at all. As usual with EFTs as long as the couplings are small enough the calculations can be trusted. This is what flux compactifications allow. If we perform a calculation assuming weak couplings and the result of the stabilisation gives us values
\begin{equation}
\mathcal{V}^{-1}\ll 1, \qquad  g_s\ll 1 \:,
\end{equation} 
then we should be able to trust the approximation. Otherwise we should discard the solution. Note also that even though string theory is not fully understood at truly strong coupling (e.g. $g_s\sim 1$), nature has been kind enough to prefer at high energies that the gauge couplings of the Standard Model tend towards a quasi-unification at weak coupling. It is then desirable that this weak coupling should be achieved in a fundamental theory, not only for computational control but also to fit with the phenomenological requirements.

\item{\it Non explicit models}. The KKLT scenario includes several non-trivial ingredients that are present in type IIB compactifications: Calabi-Yau orientifolds, three-form fluxes, warped throats, non-perturbative effects and antibranes. Each component of the scenario is possible within type IIB string theory. However, putting all together in a single model achieving moduli stabilisation in regimes where all approximations are valid is very challenging. This is what motivated us to have the simplest explicit construction that includes all the ingredients and achieves the target, that means all moduli stabilised at a de Sitter minimum (see Section~\ref{Sec:ExplicitModel}). The construction in \cite{Garcia-Etxebarria:2015lif} was already  explicit but there was no attempt to stabilise the moduli at values where the EFT approximations are trusted. This is the task we will take here.

\item{\it Runaway and non-perturbative effects}. The KKLT scenario includes fluxes that by themselves would break supersymmetry in the sense that $W_0\neq 0$. Performing the calculations of the non-perturbative corrections to $W$, knowing that beyond tree-level there is no proper perturbative minimum, was questioned \cite{Sethi:2017phn}. However as emphasised in 
\cite{Cicoli:2018kdo}, the EFT can properly be trusted if all corrections, perturbative and non-perturbative,  are assumed to be present and there is no need to start from a vacuum to perform the non-perturbative calculations 
if at the end of the calculation there is a well defined  minimum. This would have been relevant if the minimum lay in a regime where the approximations are not valid. Furthermore, in \cite{Kachru:2018aqn} several examples were considered to make the case. A general discussion of how EFTs can properly describe time dependent configurations, as the runaway discussed here, can be found in \cite{Burgess:2017ytm}.    

\item{\it Antibranes and singularities}. Some concrete solutions with several antibranes tend to give rise to singularities that may destabilise the uplift mechanism. This was the main source of criticism for the antibrane uplift for several years. However, 
at least for the case of a single probe antibrane, there is now consensus that the KKLT uplift is safe from this problem (see for instance \cite{Polchinski:2015bea, Blaback:2019ucp, Bena:2017uuz}). 

More recently a new kind of potential singularity, named {\it bulk singularity}, was identified \cite{Gao:2020xqh} (see also \cite{Moritz:2017xto}). This corresponds to the fact that the picture of a CY with warped throat is consistent if the size $R_{\rm throat}$ of the region that is warped is smaller than the radius $R_{\rm CY}$ of the CY. For KKLT in the regime of validity of the approximations, this condition cannot be satisfied and this implies singularities in the bulk. However, as stressed in \cite{Gao:2020xqh}, in LVS this problem is absent, essentially because the exponentially large volume automatically makes $R_{\rm CY}\gg R_{\rm throat}$.
To be more explicit, the claim of \cite{Gao:2020xqh}  is that the uplifting term has to be of the same order as the value of the potential at the minimum. For LVS this implies
$e^{4A_0}/\vo^{4/3}\sim W_0^2/\vo^3$. As we mentioned above,  strong warping implies  $e^{-4A_0}\gg \vo^{2/3}$. This condition corresponds to $\vo\gg W_0^2$ which is always satisfied in LVS \cite{Cicoli:2013swa}.

\item{\it 10D Formulation of non-perturbative effects in presence of antibranes}. Gaugino condensation is a 4D dynamical mechanism which should be treated in the 4D EFT since gauginos condense due to a strong coupling effect in 4D and its proper treatment depends on the logarithmic running of the gauge couplings in 4D (see for instance \cite{Burgess:1995aa} for a general discussion). However attempts to reproduce the same results from a 10D perspective are also welcome. A first attempt indicated a potential problem for the 10D case to reproduce the 4D results and was not allowing dS solutions \cite{Moritz:2017xto}. However further analysis have made it clear that a proper 10D description should reproduce the 4D results \cite{Hamada:2018qef, Carta:2019rhx, Hamada:2019ack, Kachru:2019dvo}. 

\item{\it Runaway complex structure modulus}. As mentioned in the introduction, the fact that complex structure moduli give rise to warped throats, that can be used to tune the value of the potential at the minimum,  may also destabilise the direction in complex structure moduli space and give rise again to a runaway behaviour \cite{Bena:2018fqc,Blumenhagen:2019qcg,Dudas:2019pls}.  We will see next that this is a valid issue that needs to be addressed model by model. However we emphasise that the condition $g_sM^2\gtrsim 47$ found in \cite{Bena:2018fqc} is typically satisfied if the condition for the validity of the supergravity approximation $g_sM\gg 1$ is fulfilled. We will be more explicit next regarding this important issue.

\item{\it Validity of the EFT}. As mentioned before, the validity of  the 4D EFT needs  several assumptions. In KKLT the stabilisation is usually presented in several stages. First the complex structure moduli $Z_\alpha$ and dilaton $S$ are stabilised by $D_{Z_\alpha}W=D_SW=0$ with  masses of order $m_{Z_\alpha}\sim m_S\sim B/\mathcal{V}$ with $B$ a function of the vev's of these fields. For $B\sim {\mathcal{O}}(1)$ these masses are  hierarchically larger than the gravitino mass $m_{3/2}\sim W_0/\mathcal{V}$ for $W_0 \ll 1$. However if $B \ll 1$ as it may happen if there is large warping, at least some of these fields may survive at low energies and should not be integrated out. 

In LVS the situation is different. We may consider it as moduli stabilisation in one single stage, but organising the calculation of the minimum of the scalar potential in powers of the small parameter $1/\mathcal{V}$. If at the end of the calculation the minimum is at $\mathcal{V} \gg 1$ the approximation is justified. 
For a D3 brane on a highly warped region ($e^{-4A}\gg \vo^{2/3}$) the scalar potential can be written schematically as:
\begin{equation}\label{eq:scalarP}
	V= \underbrace{V_{c.s.}}_{\mathcal{O}(1/\vo^2)}\, + \, \, \underbrace{V_{LVS}}_{\mathcal{O}(1/\vo^3)}\, \, \, + \underbrace{V_{\overline{D3}}}_{\mathcal{O}(e^{4A}/\vo^{4/3})}
\end{equation}
 Therefore in an expansion in $1/\vo$, the positive definite first term dominates and determines the extrema for the complex structure and dilaton. The warping in the last term is adjusted to make it of the same order of the second term $e^{-4A}\sim \vo^{5/3}$ to provide the dS uplift. Otherwise it could give rise to a runaway in the K\"ahler moduli directions or be subdominant. Note that here we do not have to integrate out the dilaton and complex structure moduli since the potential is extremised at each order in $1/\vo$. There may be complex structure moduli that after the process of stabilisation are lighter than most K\"ahler moduli. If we are interested on the couplings in the EFT at low energies, these fields should be included\footnote{In the presence of warping there are several regimes that should be considered depending on the values of the warp string scale $M_s^w$ and the warped Kaluza-Klein scale $M_{KK}^w$. If we want to consider couplings among moduli fields heavier than one of these scales the effective field theory should include  a finite number of KK states or even some string states. See for instance \cite{Burgess:2006mn} for a detailed discussion of all the different regimes.}. However, if we are interested in determining the vacuum, the procedure just outlined covers all the moduli fields independent of their masses.
 
As we will discuss in the next section, if the \emph{strongly} warped throat can be approximated by a Klebanov-Strassler (KS) throat \cite{Klebanov:2000hb}, then the complex structure modulus that deforms the conifold singularity has a peculiar K\"ahler potential \cite{Douglas:2007tu}. This implies a potential term for this modulus of order $\mathcal{O}(e^{4A}/\vo^{4/3})$ instead of $\mathcal{O}(1/\vo^{2})$. For this reason, as done in \cite{Bena:2018fqc,Blumenhagen:2019qcg} for KKLT, we will include this complex structure field in the stabilisation at next order in $1/\vol$ expansion. However,  we will see that the results we obtain are the same as if we had kept such modulus fixed at the value stabilised by $V_{c.s.}$. 

\end{itemize}
 It is clear that after surviving these criticisms the scenario can be considered more robust than before.   In the rest of this article we will concentrate on the second and last points that we consider the main open questions to trust antibrane uplift. We will insist to have only the goldstino as massless state coming from the antibrane in order to justify the use of the nilpotent superfield which  captures the nonlinearly realised nature of the supersymmetry broken by the antibrane. We will concentrate on the simplest realisation and will not worry  about other sectors of the model. In particular we will not attempt to construct a model with chiral observable matter. This is beyond the scope of this article and we will leave it for the  future.

\section{The dS LVS minima of the scalar potential}\label{sec:2steps}

\subsection{The scalar potential}

As we have seen, the complex structure moduli are typically fixed at higher scale than the K\"ahler moduli. This allows to study K\"ahler moduli stabilisation keeping the complex structure moduli at their fixed value. 

On the other hand, in \cite{Bena:2018fqc,Blumenhagen:2019qcg,Dudas:2019pls} it was observed that when a long warped throat is generated, the complex structure that deforms the conifold singularity is stabilised at a scale much smaller than the bulk complex structure moduli. In particular both the flux potential for $Z$ and the uplift term scale like $\left(		\frac{\zeta}{\vol}\right)^{4/3}$ around the minimum, where $\zeta\simeq e^{-\frac{2\pi K}{g_s M}}$ is the real part of the stabilised $Z$.
This means that in principle one needs to stabilise this modulus together with the K\"ahler moduli.

We now write the scalar potential for the complex structure modulus $Z$ and the K\"ahler moduli, in presence of a warped throat. The effective action for warped string compactifications was studied in \cite{DeWolfe:2002nn,deAlwis:2003sn,Giddings:2005ff,Frey:2006wv,Douglas:2007tu,Shiu:2008ry,Martucci:2009sf,Martucci:2014ska,Martucci:2016pzt}.
After stabilising the other complex structure moduli and the axio-dilaton, one is left with an effective  \K potential that includes also the contribution for $Z$, that is valid in the strongly warped regime \cite{Douglas:2007tu,Blumenhagen:2019qcg,Dudas:2019pls}:
\begin{equation}\label{eq:Kpotential}
	K = K_{\rm LVS}+K_X+\frac{c' \xi' |Z|^{\frac{2}{3}}}{\vol^\frac{2}{3}}\:,
\end{equation}
where $\xi'=9g_s M^2$ and $c'$ is an order one numerical factor whose value was computed to be $c'\simeq 1.18$ \cite{Douglas:2007tu}. 
The effective superpotential is
\begin{equation}\label{eq:totalW}
	W = W_{0}-\frac{M}{2 \pi i}Z (\log{Z}-1)+i K s Z + A_s e^{-aT_s} + \mu(Z) X \:,
\end{equation}
where $W_0$ is the contribution of all the other complex structure moduli, stabilised at higher energies;\footnote{Typically what is called $W_0$ includes the contribution of all complex structure and then it would be the sum of the first two terms on the r.h.s. of \eqref{eq:totalW}. However, as it will be clear in the following, the second term will contribute negligibly to the complex structure superpotential.
} 
$M \text{ and } K$ are the quantised $F_3$-form flux on the $S^3$ and $H_3$-form flux on the dual 3-cycle, respectively; $s=\frac{1}{g_s}$ is the dilaton, stabilised at higher energies.

After using the approximation $\vol\gg 1$,
 the sugra scalar potential corresponding to the K\"ahler potential \eqref{eq:Kpotential} and the superpotential \eqref{eq:totalW} is\footnote{In principle $\eta$ depends on $Z$. However, terms including derivatives of $\eta$ with respect to $Z$ are set to zero once one imposes $X=0$ in the vacuum. So effectively $\mu$ can be taken as constant.} 
\begin{equation}\label{eq:V}
	\begin{split}
	V_{tot}=&\frac{8 a^2 A^2 g_s \sqrt{\tau_s} e^{-2 a \tau_s}}{3 \kappa_s \vol}+\frac{4 a A g_s \tau_s e^{-a \tau_s}}{\vol^2} \left( W_{0} \cos (a \theta_s+\phi )+\zeta \frac{M}{2\pi} \sin (a \theta_s+\sigma )\right)\\
		     &+\frac{\zeta ^{4/3}}{c' M^2 \vol^{4/3}} \left[\frac{c' c''}{\pi g_s}+\frac{M^2 \sigma ^2}{4\pi^2}+\left(\frac{M}{2\pi}\log\zeta+\frac{K}{g_s}\right)^2\right]\\
		     &+\frac{3 \xi}{2  \sqrt{g_s} \vol^3} \left[W_{0}^2-16\zeta^2\left(\frac{M}{2\pi}\log\zeta+\frac{K}{g_s}\right)^2 +\frac{M^2\zeta^2}{\pi^2}\left(\frac{1}{4}+\frac{2\pi K}{M g_s}-4\sigma^2+\log\zeta\right) \right.\\
		     &\left. + 2\frac{M}{\pi}W_{0}\zeta\sigma \cos(\sigma-\phi)+W_{0}\zeta\sin(\sigma-\phi)\left(\frac{M}{\pi}+2\frac{M}{\pi}\log\zeta+\frac{4K}{g_s}\right)\right] \:.
	\end{split}
\end{equation}
Here, with abuse of notation, we have replaced $W_0\rightarrow W_0 e^{i\phi}$, where now $W_0$ is real and positive and $\phi$ is the phase of the complex $W_0$ appearing in \eqref{eq:totalW}.

\subsection{dS minima}

We are interested in minima where the warp factor takes large values. This allows to reduce the uplifting term so that it compensate the negative AdS vacuum energy of the LVS minimum and give a tiny dS minimum.  The LVS AdS minimum is at $V_{\rm LVS}\sim \frac{W_0^2}{\vol^3}$, while the uplifting term is of the order $V_{\rm uplift}\sim \frac{\zeta^{4/3}}{\vol^{4/3}}$ (see equation \eqref{uplifttermDudasEtAl}). Hence, to obtain a small dS minimum we will work in the approximation where 
		\begin{equation}\label{eq:zetaSmall}
			\zeta^{4/3}	\sim \frac{W_0^2}{\vol^{5/3}} \:.
		\end{equation}
This automatically imposes $\zeta\ll 1$ in a LVS regime. 

We now approximate the potential \eqref{eq:V} under the assumption \eqref{eq:zetaSmall} and compute its first derivatives to find the minima. 
We  obtain:
\begin{equation}
	\begin{split}
	V_{tot}\simeq&\frac{8 a^2 A^2 g_s \sqrt{\tau_s} e^{-2 a \tau_s}}{3 \kappa_s \vol}+\frac{4 a A g_s \tau_s e^{-a \tau_s} W_{0}}{\vol^2} \cos (a \theta_s+\phi )\\
		     &+\frac{\zeta ^{4/3}}{c' M^2 \vol^{4/3}} \left[\frac{c' c''}{\pi g_s}+\frac{M^2 \sigma ^2}{4\pi^2}+\left(\frac{M}{2\pi}\log\zeta+\frac{K}{g_s}\right)^2\right]+\frac{3 \xi}{2  \sqrt{g_s} \vol^3} W_{0}^2
	\end{split}
\end{equation}

We note that $\theta_s$ and $\sigma$ are stabilised at $\theta_{s_0}=\frac{\pi-\phi}{a}$ and $\sigma_0=0$. Moreover one sees that these moduli are decoupled from $\zeta,\tau_b,\tau_s$, since:
$$
\frac{\partial^2 V_{tot}}{\partial x_i\partial y_j}|_{\sigma_0, \theta_{s_0}}=0\,,
$$
where $x_i=\{\theta_s, \sigma\}$ and $y_j=\{\tau_s, \tau_b, \zeta\}$.\footnote{
Eventually we will check that the Hessian is in fact positive definite for these values.}

Fixing $\sigma$ and $\theta_s$ at their minimum values, the  scalar potential for the remaining three moduli becomes 
\begin{eqnarray}\label{eq:Vappr}
	V&=&\frac{8 a^2 A^2 g_s \sqrt{\tau_s} e^{-2 a \tau_s}}{3 \kappa_s\mathcal{V}}-\frac{4 a A g_s \tau_s W_0 e^{-a \tau_s}}{\mathcal{V}^2}+\frac{3 W_0^2 \xi}{2 \sqrt{g_s}\vol^3} +  \nonumber \\  \\
	&& +\frac{3 \zeta ^{4/3}}{8 \pi^2 c' \vol^{4/3}} \left[\frac{8\pi c' c''}{3g_s M^2}+\frac{8\pi^2 K^2}{3 g_s^2 M^2}+\frac{8\pi K}{3 g_s M}\log\zeta+\frac{2}{3}\log^2\zeta\right] \:. \nonumber 
\end{eqnarray}
The resulting  scalar potential, at this level of approximation, is given by the usual LVS one plus an uplift term.
 
By computing the first derivatives of \eqref{eq:Vappr}, we get the values of the moduli at the minimum:
\begin{subequations}
	\begin{align}
		\partial_\zeta V &=0 \Leftrightarrow \zeta = e^{-\frac{2\pi K}{g_s M}-\frac{3}{4}+\sqrt{\frac{9}{16}-\frac{4\pi}{g_s M^2}c' c''}}\:,\label{eq:zMin}\\
		\partial_{\tau_s}V &=0 \Leftrightarrow  \tau_b^{3 / 2}=\frac{3 e^{a \tau_{s}} W_{0} \kappa_s\sqrt{\tau_{s}}}{a A}\frac{\left(1-a \tau_{s}\right)}{\left(1-4 a \tau_{s}\right)}\:,\label{eq:volMin}\\
		\partial_{\tau_{b}} V &=0 \Leftrightarrow \tau_{s}^{3/2}\frac{16a\tau_s(a \tau_s-1)}{(1-4 a \tau_s)^2}=\frac{\xi}{g_s^{3/2}\kappa_s}+\frac{8q_0\zeta^{4/3}\tau_b^{5/2}}{27 g_s \kappa_s W_0^2}\:,
		\label{eq:TauSMin}
	\end{align}
\end{subequations}
where $q_0 \equiv \frac{3}{8\pi^2 c'}\left(\frac{3}{4}-\sqrt{\frac{9}{16}-\frac{4\pi c' c''}{g_s M^2}}\right)$ 
and the relations \eqref{eq:zMin} and \eqref{eq:volMin} were used in~\eqref{eq:TauSMin}.
We have moreover approximated $\vol\simeq 	\tau_b^{3/2}$.
Again we use the fact that $a\tau_s\gg 1$ (that is also necessary to have large volume) to approximate the  equations \eqref{eq:zMin}-\eqref{eq:TauSMin} to:
\begin{subequations}
	\begin{align}
		\partial_\zeta V &=0 \Leftrightarrow \zeta = e^{-\frac{2\pi K}{g_s M}-\frac{3}{4}+\sqrt{\frac{9}{16}-\frac{4\pi}{g_s M^2}c' c''}}\:,\label{eq:zMinSimp}\\
		\partial_{\tau_s}V &=0 \Leftrightarrow  \tau_b^{3 / 2}\simeq \frac{3  W_{0} \kappa_s \sqrt{\tau_{s}}}{4a A}\,e^{a \tau_{s}}\:,\label{eq:volMinSimp}\\
		\partial_{\tau_{b}} V &=0 \Leftrightarrow \tau_{s}^{3/2}\simeq \frac{\xi}{g_s^{3/2}\kappa_s}+\frac{8\rho\,\tau_b^{5/2}}{27 g_s  \kappa_s W_0^2}\:,\label{eq:TauSMinSimp}
	\end{align}
\end{subequations}
where we have defined  $\rho\equiv q_0\zeta^{4/3}$, with $\zeta$ given by the first equation, i.e.
\begin{equation}\label{eq:rhoDefinition}
\rho \equiv q_0' e^{ -\frac{8\pi K}{3g_sM} } \qquad\mbox{with}\qquad q_0'\equiv q_0 e^{-\frac{32\pi^2c'}{9}q_0}    \:.
\end{equation} 
For later convenience, notice that $q_0'$ does not depend on the flux number $K$.

We observe that the expression obtained for $\zeta$ is compatible with the result in \cite{Bena:2018fqc,Blumenhagen:2019qcg,Dudas:2019pls}.
Note that $\zeta$ is shifted with respect to the KS vacuum, where it was $\zeta=e^{-\frac{2\pi K}{g_s M}}$. The important fact is not this shift, that is irrelevant,\footnote{At worst, it scales $\zeta$  by a factor close to $1$.} but the fact that the minimum exists only if the argument of the square root is positive. This happens if \cite{Dudas:2019pls}
\be\label{benabound}
g_sM^2 \gtrsim 47.
\ee
Differently, there is a runaway of the scalar potential \cite{Bena:2018fqc}.

The approximated equations \eqref{eq:zMin}-\eqref{eq:TauSMin} give the same results that were found in \cite{Aparicio:2015psl}: in that paper, the authors considered the potential of the two K\"ahler moduli where the value of $\zeta$ was assumed to be fixed at higher scales (at the KS solution). We conclude that even if the modulus $Z$ has a small mass, its stabilisation is however decoupled form the K\"ahler moduli one.

Finally, we evaluate the potential at the minimum:
\begin{equation}\label{eq:vMin}
	V_{min} = \frac{5 q_0 \zeta^{4/3}}{9 \tau_b^2}-\frac{12 W_0^2 g_s \kappa_s\tau_s^{3/2}}{\tau_b^{9/2}}\frac{(a \tau_s-1)}{(1-4a\tau_s)^2} \simeq \frac{5 \rho}{9 \tau_b^2}-\frac{3 W_0^2 g_s \kappa_s\tau_s^{1/2}}{4a\,\tau_b^{9/2}}\:,
\end{equation}
that in the limit \eqref{eq:zetaSmall} under consideration is approximately Minkowski.  Depending on the value of $\rho$, the minimum can be AdS, Minkowski or dS.

\subsection{Moduli masses}

One  can compute the masses for the moduli at the minimum, by looking at the eigenvalues of the matrix $\frac12 \mathrm{K}^{-1} \partial^2V$, where $ \partial^2V$ is the Hessian of the scalar potential and $\mathrm{K}^{-1}$ is the inverse of the K\"ahler metric on the scalar field space. We consider the $3\times 3$ block along the three moduli $\tau_s,\tau_b,\zeta$ (the phase modes are decoupled as we have seen above).

The characteristic polynomial of the $3\times 3$ matrix is a third degree polynomial of the form $-\lambda^3+b\lambda^2+c\lambda+d=0$, where at leading order in our approximation (in particular $a\tau_s\gg 1$) we have
\begin{equation}
b \simeq A + B\,, \qquad c \simeq - A\, B \,, \qquad d \simeq A B C \:,
\end{equation}
with
\begin{eqnarray}
A &\equiv& \frac{2 q_0^{1/2} \rho^{1/2}}{9 g_s c' M^2 \tau_b} \:,\\
B &\equiv& \frac{g_s W_0^2 a^2 \tau_s^2}{\tau_b^3} \:,\\
C &\equiv&\frac{3}{2\tau_b^2} \left(\frac54 \, \frac{27 g_sW_0^2\kappa_s\tau_s^{1/2}}{20a\tau_b^{5/2}} - \rho\right) \:.
\end{eqnarray}
Notice that $A,B,C	\ll 1$ and hence $b\gg c \gg d$. In particular $C\ll A,B$, which implies $\frac{d}{bc}\ll 1$.
One can then solve the cubic equation perturbatively and obtain
\begin{equation}\label{eq:modulimasses}
  m_1^2 \simeq b = A+B\,, \qquad m_2^2 \simeq -\frac{c}{b} = \frac{A\,B}{A+B} \,, \qquad m_3^2 \simeq -\frac{d}{c}=C\:.
\end{equation}
For the typical values of the parameters, $B$ is bigger than $A$. When this happens, we have $m_1^2\simeq B$ and $m_2^2\simeq A$, i.e. $m_1$ is equal to the mass of the modulus $\tau_s$ in LVS. However what is most important is that $m_3$ is much smaller than the other two masses. Moreover, it is not always positive (we will see the consequences in the next section).

\subsection{Bounds on the warp factor}

Let us keep the parameters $W_0,A,\kappa_s,a,g_s$ fixed. The value of $\rho$ changes with different choices of the fluxes $M,K$. As a matter of fact, one can obtain a de Sitter minimum only for a limited range of values of $\rho$ when we fix the other parameters. 

One has Minkowski when $V_{min}=0$, i.e. when $\rho$ is equal to
\begin{equation}\label{eq:rhoMink}
		\rho_{\rm low} \simeq \frac{27 g_s W_0^2 \kappa_s\tau_s^{1/2}}{20a\,\tau_b^{5/2}}\:,
	\end{equation}
where $\tau_s$ and $\tau_b$ are functions of the parameters $W_0,A,\kappa_s,a,g_s$, that are determined by solving the equations \eqref{eq:zMin}-\eqref{eq:TauSMin}.

Increasing the value of $\rho$,   $m_3^2$ in \eqref{eq:modulimasses} becomes negative at some point,
and the solution of \eqref{eq:zMin}-\eqref{eq:TauSMin} is not a minimum of the scalar potential anymore. The value of $\rho$ when this happens~is
\begin{equation}\label{eq:rhoMax}
		\rho_{\rm up} \simeq \frac{5}{4}	\,\frac{27 g_s W_0^2 \kappa_s\tau_s^{1/2}}{20a\,\tau_b^{5/2}} \:.
	\end{equation}

We see that the order of magnitude of admissible $\rho$ (hence of the warp factor) is fixed by the other parameters in the game, i.e. $W_0,A,\kappa_s,a,g_s$. In particular, on a dS minimum $\rho$ will be given by
\begin{equation}\label{eq:rhoalpha}
		\rho \simeq \alpha	\,\frac{27 g_s W_0^2 \kappa_s\tau_s^{1/2}}{20a\,\tau_b^{5/2}}
	\end{equation}
where $\alpha$ is a number in $]1,\frac{5}{4}[$.


Using this fact, we can rewrite the equation \eqref{eq:TauSMinSimp} as
\begin{equation}\label{eq:tausApprox}
 \tau_{s}^{3/2}\simeq \frac{\xi}{g_s^{3/2}\kappa_s}+\frac{2\alpha \,\tau_s^{1/2}}{5a}  \:.
\end{equation}
Solving the cubic equation in $\tau_s^{1/2}$ one realises that $\tau_s$ is shifted with respect to the LVS minimum ($\tau_s\simeq \frac{\xi^{2/3}}{g_s\kappa_s^{2/3}}$) by the small quantity $\delta\tau_s=\frac{4\alpha }{15a}\lesssim 1$. This small shift affects the exponentially large $\tau_b$ by a factor of $\mathcal{O}(1)$.
We have then shown that the uplift term in the potential modifies the relations determining the AdS LVS minimum only at subleading order.

\section{Consistency conditions and limits on D3 tadpole}\label{sec:boundsond3tadpole}

As we have said above, the 3-form fluxes $M$ and $K$ contribute to the D3-charge, by the positive number $MK$. 
In this section we analyse what is the minimal value of $MK$ that is allowed within the approximation used to obtain the minima. We will see that this number is typically large, even larger than what expected by the no-runaway condition found in \cite{Bena:2018fqc,Dudas:2019pls}.

\subsection{Constraints}\label{Sec:Constraints}

To find a dS minimum, one needs to solve equations \eqref{eq:zMin}-\eqref{eq:TauSMin} (and check that the solution is actually a minimum). 
These equations are valid only in some regions of the moduli space ($\tau_b,\tau_s,\zeta$) and hence of the parameter space ($W_0,g_s,a,A,M,K,\chi(X)$). 

We now summarise the conditions that define these regions of validity (some of which we already mentioned before). We will say explicitly when the condition is automatically satisfied by any solution of the equations \eqref{eq:zMin}-\eqref{eq:TauSMin} and when this puts constraints on the range of the parameters.
\begin{itemize}
	\item Since we want to obtain a de Sitter minimum, the potential must be stabilised at a positive value. The potential at the minimum should be very small:  
		\begin{equation}\label{eq:MinkdS}
			 V_{\min }\gtrsim 0
		\end{equation}
	\item The minimum should be found in a regime where one keeps control over the EFT. 
	First, the K\"ahler moduli should be stabilised in a region with large volume (so that we can trust the $\alpha'$ expansion\footnote{Notice that this condition has to be imposed on the string-frame compactification volume ($\vol_s=\vol_E g_s^ {3/2}$) rather than on the Einstein-frame one (in which we are working).}):
			\begin{equation}\label{eq:largeV}
				\vol\gg\frac{\xi}{g_s^{3/2}}\gg1 \qquad\Longleftrightarrow\qquad \tau_b\gg\tau_s \:.
			\end{equation}
This condition guarantees that the  \K moduli are stabilised within the \K cone in the class of models we consider (where the \K cone condition reads $\tau_b>\tau_s$).

Secondly, one requires negligible multi-instanton effects in the non-perturbative corrections, i.e. $a \tau_s\gg1$. In particular we will require $\tau_s\gg 1$ in order to be consistent with the supergravity approximation. 

In the minima of our potential, all these conditions are satisfied once one imposes $g_s\ll 1$, that is anyway necessary to exclude deviations from the perturbative string approximation. The parameter $a$ may actually be very small, if the non-perturbative effect is generated by a condensing group with high rank. However, as explained in Section~\ref{Section:LVSmoduliStab}, the only condensing group that we  consider is $SO(8)$, that has $a=\frac{\pi}{3}$. Hence in our analysis $a\gtrsim 1$.
	
\item The low-energy  supergravity provides a valid 4d description only if the following hierarchy holds:
		\[M_p\gg M_s\gtrsim M_{KK}^{(i)}\gg m_{moduli}, m_{3/2}\]
		The requirement $M_{KK}\gg m_{3/2}$, in particular, can be expressed in terms of the parameters of the model as \cite{Cicoli:2013swa,AbdusSalam:2020ywo}:
		\begin{equation}\label{eq:w0Max}
			\sqrt{\frac{\kappa}{\pi}}W_0\ll\vol^{1/3}\:,
		\end{equation}
 where\footnote{Our $W_0$ is rescaled by a factor of $\sqrt{4\pi}$ w.r.t. what is found in \cite{AbdusSalam:2020ywo}.}   $\kappa=\frac{g_s e^{K_{cs}}}{2}$.
		This puts an upper bound on $W_0$ \cite{AbdusSalam:2020ywo}.
		
\item 
We want a highly warped throat, i.e.
		\begin{align}
		 	 \vol^{2/3} \rho \ll1  
		\end{align}
		where $\rho\equiv q_0\zeta^{4/3}$, with $\zeta$ evaluated at the minimum of the potential. 
		Moreover, we need to request:
		\begin{equation} \label{eq:stringModes}
			\rho^{1/4}\vol^{2/3}\gg\frac{\sqrt{\kappa}W_0}{g_s^{1/4}\pi^{1/2}}\sim 1
		\end{equation}
		so that the massive string states of the $\overline{D3}$-brane at the tip of the throat, which are redshifted to lower masses, are still negligible with respect to $m_{3/2}$. 
		
		These two conditions are always satisfied at the dS minima of the potential.  This is clear from using the  equations \eqref{eq:zMinSimp}-\eqref{eq:TauSMinSimp} and taking
		the expression  \eqref{eq:rhoalpha} of $\rho$ at the minimum.

	\item The size of the $S^3$ at the tip of the conifold ($R^2_{S^3}\sim \alpha' g_s M$) should stay larger than the string length in order to trust the KS solution, hence:
			\begin{equation}\label{eq:gSm}
				g_s M\gg1.
			\end{equation}
			For fixed (small) $g_s$ this puts a lower bound on the flux $M$. This flux appears in the warp factor \eqref{eq:warpFactorTip}: if one wants to generate strong warping (and hence a $\rho$ like in \eqref{eq:rhoalpha}), one needs that the exponent is not too small; hence, if $M$ is large, $K$ should be large as well and the flux contribution to the D3-charge may be very large. 


	\item We want to to avoid the runaway discussed in \cite{Bena:2018fqc,Dudas:2019pls}. Hence we need 
		\begin{equation}\label{eq:gSm2}
			g_s M^2\gtrsim 47
		\end{equation}		
We notice that this constraint is less strong than \eqref{eq:gSm}. Hence fluxes satisfying that condition  do not typically have the runaway problem.

\item Finally, as pointed out in \cite{Dudas:2019pls,Gao:2020xqh}, in the KS solution the $Z$ field is not a modulus, it being fixed to its supersymmetric value. Taking an off-shell $Z$-dependence of the warp factor is trustable only if the stabilised value of $Z$ does not deviate much from the flux stabilised value. For the models under consideration this condition is fulfilled as $\zeta$ in \eqref{eq:zMin} differ from the KS value by a factor of order one (as it can be checked in our explicit model in Section~\ref{Sec:ExplicitModel}).

\end{itemize}

\subsection{Bounds on the flux D3-charge}\label{Sec:conditions1-3}

From Section~\ref{Sec:Constraints} we  conclude the following. Take a dS minimum of the scalar potential, satisfying then \eqref{eq:zMinSimp}-\eqref{eq:TauSMinSimp} and \eqref{eq:rhoalpha}. To be consistent with all the approximations we used, it is enough that the parameters of the model  fulfill the following three conditions: 
\begin{enumerate}
\item[(1)] $g_s\ll 1$ such that $\vol\gtrsim 10^4$;
\item[(2)] $\sqrt{\frac{\kappa}{\pi}}W_0\ll\vol^{1/3}$ (we will take $\mathcal{O}(10)$ as limiting ratio);
\item[(3)] $g_s |M| \gg 1$, we will take $g_s |M| \gtrsim 5$.\footnote{We assume that  this is enough for the analysis of \cite{Kachru:2002gs} on the \ad3 stability to hold. In particular, with this constraint one always finds $M>12$ in the perturbative regime.}
\end{enumerate}

In the following we will show that the conditions (1)-(3) above, put a \emph{large} lower bound for the flux D3-charge, as has been recently observed also in \cite{Demirtas:2020ffz}. This large positive D3-charge is typically difficult to cancel in a perturbative type IIB setup, even though in our opinion the situation is not as bad as suggested in \cite{Bena:2018fqc}.
This shows that this uplift mechanism is to be taken carefully, at least when one wants to use the KS approximation.

The maximal value of $g_s$ that satisfies condition (1) strongly depends on the Euler characteristic $\chi(X) $ of the CY threefold, on $\kappa_s$ and on the value of $a$ that appears in the non-perturbative superpotential. This can be understood substituting equation  \eqref{eq:TauSMinSimp} (or its approximation \eqref{eq:tausApprox}) into equation \eqref{eq:volMinSimp}.

For a fixed volume $\vol$, one can increase the value of $g_s$ 
by making $W_0$ larger. However, condition~(2) puts an upper bound on possible choices for $W_0$. 

Once the value of $g_s$ is chosen, condition (3) generates a bound on $M$.
In order to minimise the D3-charge $MK$, one should take $M$ close to this bound.  Since the bound is smaller for larger $g_s$, these models typically prefer the biggest $g_s$ compatible with condition~(1).
Since the warp factor $\rho$, defined in \eqref{eq:rhoDefinition},
can only vary in a small range (corresponding to $1\leq \alpha<\frac{5}{4}$, see \eqref{eq:rhoalpha}), for a fixed  $M$, $K$ can take few values. To obtain the minimal D3-charge $MK$ of the fluxes in the throat one takes the lowest $M$ and then choose the smallest among the associated $K$.

There is a caveat: if (even at the largest value of $g_s$ compatible with perturbation theory) one takes $a$ and $\chi(X)$ large or $\kappa_s$ small, this might produce a very large value for $\vol$. This may seem good; however a very large volume means a very small value of $\rho$ (see equation \eqref{eq:rhoalpha}). In order to have this, one needs to take a large ratio $\frac{K}{g_sM}$ (see equation \eqref{eq:zMinSimp}), that implies a large $K$, even for  $g_sM$ not so big. This eventually leads to a large D3-charge $MK$.


In general, one notices that the values of the parameters that minimise the tadpole $
MK$ are typically at the boundary of our consistency conditions.
In order to strengthen these considerations,
we made a  rough scan over  models with fixed $\chi(X)$, $a$ and $\kappa_s$,
by varying the parameters $g_s,W_0,K,M$ (we keep $A=1$). 
We see which ones allow to find a dS minima that satisfy conditions (1)-(3) and at the same time keep the D3-charge $MK$ of the throat fluxes to be small.

We proceed in the following way: for each value of $W_0$ in the range $[1,30]$ (with step $1$) and $g_s$ in $[0.01, 0.3]$ (with step $0.001$), we compute the lower and upper bounds for the warp factor $\rho$ given in \eqref{eq:rhoMink} and \eqref{eq:rhoMax}. We then select $M$ and $K$ such that $\rho\in[\rho_{\rm low},\rho_{\rm up}]$ and the conditions (1)-(3) are fulfilled. This allows to select the minimal $MK$ for that point in the $(g_s,W_0)$-space. Scanning over all $(g_s,W_0)$ that give minima compatible with conditions (1)-(2), we obtain the minimum for $MK$ for  given $\chi(X)$, $a$ and $\kappa_s$. Notice that once we choose $X$, the parameters
$\chi(X)$, $a$ and $\kappa_s$ are basically fixed.\footnote{Typically, for a given CY $X$ and a choice of orientifold involution, there are few possible non-perturbative divisors. Moreover the involution already says if they will support instantonic D3-branes or gaugino condensation on an $SO(8)$ D7-brane stack.} We leave the scan over the CY's parameters $\chi(X),a,\kappa_s$ for a future work. However, in order to have an idea how the results change with them, we make the scan over $g_s,W_0$ for  few different values of $\chi(X)$, $a$, $\kappa_s$.

As we said, we distinguish between two values of $a$, i.e. $a=\frac{\pi}{3}$ when the non-perturbative superpotential is generated by gaugino condensation, and $a=2\pi$ when it  is given by a D3-instanton. For $a=\frac{\pi}{3}$ the lowest values of $MK$ are given, for  few different choices of $\kappa_s$  and $\chi(X)$, by (we report also which values of $g_s,W_0$ produce each given $MK$)
	\begin{equation}\label{Tab:aPi3}\begin{array}{  |c c | c c c | }
		\cline{1-5}
			(MK)_{\{g_s,W_0\}}& &\multicolumn{3}{ c |}{-\chi(X)}\\
		\cline{3-5}
			& & 200 & 350 & 500\\
		\cline{1-5}
		\multicolumn{1}{| c  }{\multirow{4}{*}{$\kappa_s$}}
			& \multicolumn{1}{| c |}{0.05} & 96_{\{0.313, 10\}} & 160_{\{0.333, 15\}} & 224_{\{0.331, 6\}} \\
			& \multicolumn{1}{| c |}{0.1} & 110_{\{0.228, 10\}} & 102_{\{0.295, 11\}} & 128_{\{0.313, 5\}} \\
			& \multicolumn{1}{| c |}{0.5} & 290_{\{0.087, 4\}} & 200_{\{0.126, 5\}} & 160_{\{0.157, 8\}} \\
			& \multicolumn {1}{| c |}{1}   & 312_{\{0.065, 7\}} & 305_{\{0.082, 3\}} & 240_{\{0.106, 3\}}\\
		\cline{1-5}
	\end{array}\end{equation}
As we have observed above, larger values of the Euler characteristic $|\chi(X)|$ and smaller values of $\kappa_s$ provide larger volumes at fixed $g_s$. Hence, in general, one expects to find a larger $(MK)_{\rm low}$ as one increases $\kappa_s$ and/or decreases $|\chi(X)|$. This is  what we observe in \eqref{Tab:aPi3}. However, the cases $\chi=-500$ and $\kappa_s\leq 0.1$ are special because of the caveat we pointed out above about large $\chi(X)$ and small $\kappa_s$. 
The same observation can be done for $|\chi|\geq 350$ and $\kappa_s=0.05$.

The following table shows the minimum tadpoles for $a=2\pi$:
	\begin{equation}\begin{array}{ | c c | c c c | }
		\cline{1-5}
			(MK)_{\{g_s,W_0\}}& &\multicolumn{3}{ c |}{\chi}\\
		\cline{3-5}
			& & 200 & 350 & 500\\
		\cline{1-5}
		\multicolumn{1}{| c  }{\multirow{3}{*}{$\kappa_s$}}
			& \multicolumn{1}{| c |}{0.1} & 512_{\{0.325, 4\}} & 768_{\{0.332, 15\}} & - \\
			& \multicolumn{1}{| c |}{0.5} & 144_{\{0.328, 1\}} & 224_{\{0.333, 12\}} & 304_{\{0.328, 9\}} \\
			& \multicolumn {1}{| c |}{1}   & 225_{\{0.201, 4\}} & 128_{\{0.313, 7\}} & 176_{\{0.313, 9\}}\\
		\cline{1-5}
	\end{array}\end{equation}
Here, the large value of $a$ makes the caveat works for all chosen value of $\chi(X)$ and $\kappa_s$. The volume is always big in these examples. It was not possible to find a suitable model with $(\chi, \kappa_s)=(500, 0.1)$ in the explored range of parameters. Anyway, if it exists, it is expected to have $MK\gtrsim10^3$.  In general, we can deduce from this analysis that taking larger values of $a$ does not improve the results in terms of the tadpole.

We finally observe that the values we obtain are smaller than the recent lower bound found in \cite{Bena:2020xrh}, i.e. $MK\gtrsim 500$. This large value was obtained by requiring a very small warp factor, necessary in KKLT to preserve the stability of the K\"ahler moduli after uplift. In the LVS vacua we have analysed in this paper, the stability of the vacua has been taken into account by considering $\rho$ as in \eqref{eq:rhoalpha}. Hence, we claim that for LVS the bound of \cite{Bena:2020xrh} is a bit smaller: we typically find $MK\gtrsim 100$.


\section{An explicit model with \ad3-branes uplift}\label{Sec:ExplicitModel}

In this section, we present an explicit model. We will choose a CY three-fold that is an hypersurface in a toric variety. The orientifold involution will be chosen such that there will be O3-planes that collapse to each other by taking a conifold limit. This will reproduce a situation with a warped throat modeled on a deformed conifold. This has been discussed in \cite{Garcia-Etxebarria:2015lif} and it provides the simplest situation with a nilpotent superfield studied in \cite{Kallosh:2015nia}.

We will show that this model admits a dS minimum that satisfies the conditions (1)-(3) of Section~\ref{Sec:conditions1-3}. 

\subsection{Geometric setup}

We consider the toric ambient space characterised by the following weights and SR-ideal 
\begin{equation}
\begin{array}{c|cccccc|c}
 & z & u_1 & u_2 & v & w & \xi &  D_\textmd{H} \tabularnewline \hline 
    \mathbb{C}_1^* & 0  &  1  &  1  &  2  &  3  &  7  & 14\tabularnewline 
    \mathbb{C}_2^* & 1  &  0  &  0  &  0  &  1  &  2  & 4\tabularnewline
\end{array}\label{eq:model3dP8:weightm}\,,
\qquad\qquad
{\rm SR}=\{z\,w,\, u_1\, u_2\,v \, \}\,.
\end{equation}
The CY threefold $X$ is an hypersurface in this ambient space, defined by an equation with degrees given in last column of \eqref{eq:model3dP8:weightm} (it is the number 39 in the database \cite{Altman:2014bfa}).

This CY $X$ has Hodge numbers $h^{1,1}=2$ and $h^{1,2}=132$, and Euler characteristic $\chi(X)=-260$.
The divisors classes $D_z$ (with representative $\{z=0\}$) and $D_u$ (with representative $\{a_1u_1+a_2u_2=0\}$ for arbitrary $a_1,a_2\in\mathbb{C}$) make up an integral basis for $H^2(X,\mathbb{Z})$.\footnote{We use the same symbol for the 4-cycles and the Poincar\'e dual 2-forms.}
The intersection form takes the  expression
\begin{equation}
  I_3 = 9D_z^3 - 3 D_z^2D_u + D_zD_u^2\:. 
\end{equation}
The second Chern class of the Calabi-Yau is
\begin{equation}\label{c2CY2ndExample}
c_2(X) =66 D_u^2 - 8 D_z^2 \: .
\end{equation}
For what follows, it will be useful  to use $\{D_w,D_z\}$ as a basis of $H_4(X)$ (even though non-integral, i.e. it generates the integral divisors by rational linear combinations). In this basis the intersection form takes the simple form
\begin{equation}
  I_3 = 9D_w^3  + 9D_z^3\:. 
\end{equation}

Let us expand the K\"ahler form on the basis $\{D_w,D_z\}$: 
\begin{equation}
 J = t_wD_w + t_zD_z \:.
\end{equation}
We can then express the volumes of the divisors $D_w$ and $D_z$ in terms of the parameters $t_z,t_w$:
\begin{equation}
  \tau_w\equiv {\rm vol}(D_w)=\tfrac12\int_{D_w} J^2 = \frac92 t_w^2 \,,\qquad\qquad   \tau_z\equiv {\rm vol}(D_z)=\tfrac12\int_{D_z} J^2 = \frac92 t_z^2 \:.
\end{equation}
The volume of the CY three-fold is then
\begin{equation}\label{eq:volExplModel}
\vol = \tfrac16 \int_X J^3 = \frac{3}{2} \left( t_w^3+t_z^3 \right) = \frac{\sqrt{2}}{9}
\left( \tau_w^{3/2}-\tau_z^{3/2} \right)
\end{equation}
We see that the volume takes the Swiss cheese form \eqref{SCvolume}. 
We then identify $\tau_b=\frac{1}{3}\left(\frac{2}{3}\right)^{1/3} \tau_w$, $\tau_s=\tau_z$ and $\kappa_s=\frac{\sqrt{2}}{9}$.
Moreover, the divisor $D_z$ is a $\mathbb{P}^2$, i.e. a  rigid cycle with $h^{1,1}=0$. It will actually support the non-perturbative effect.

\subsection{Involution} 

We consider the involution 
\begin{equation}\label{Invol}
 \sigma\,: \qquad w \mapsto - w \:.
\end{equation}
The CY three-fold equation must be restricted to be invariant under this involution, i.e only monomials with even powers of $w$ can appear.  In particular, the defining equation turns out to be\footnote{We have reabsorbed the linear term in $\xi$ by a redefinition of $\xi$ itself.}
\begin{eqnarray}\label{CYeq2ndMd}
\xi^2 &=& w^4 \left[v+P_2(u)\right]  -2 b \, w^2 z^2\left[v^4+v^3Q_2(u)+v^2Q_4(u)+v\,Q_6(u)+Q_8(u)\right] + \\
 && + c\,z^4 \left[v^7+v^6 R_2(u)+v^5 R_4(u)+v^4 R_6(u)+v^3 R_8(u)+v^2 R_{10}(u)+v R_{12}(u)+ R_{14}(u)     \right]\nonumber
\end{eqnarray}
  where $P_n(u),\,Q_n(u),\,R_n(u)$ are polynomials of degree $n$ in the coordinates $u_1,u_2$ and $b,c\in\mathbb{C}$.

Let us consider the fixed point locus under the involution \eqref{Invol}.
It is made up of two codimension-1 loci at $w=0$ and $z=0$ and two isolated fixed points at the intersection $\xi=u_1=u_2=0$. Hence, by implementing this orientifold involution, one obtains two O7-planes in the classes $[D_{O7_1}]=D_w$ and $[D_{O7_2}]=D_z$, and two O3-planes. Notice that $D_{O7}^3=18$.

The Euler characteristics of the O7-planes divisors are given by $\chi(D)=\int_D c_2(D)$.\footnote{Since $D$ is a divisor of $X$, we can use the adjunction formula to obtain $c_2(D)=c_2(X)+c_1(D)^2$. Since $X$ is a CY, $c_1(D)=-D$. } In the present model we have $\chi(O7_1)=\chi(D_w)=75$ and $\chi(O7_2)=\chi(D_z)=3$.

We can now calculate, by means of Lefschetz fixed point theorem, the values of $h^{1,2}_-$ (that gives the number of complex structure deformations of the invariant equation \eqref{CYeq2ndMd}) and $h^{1,2}_+$ (that gives the number of abelian bulk vectors). The theorem states the following relations between the (even and odd) Betti numbers and the Euler characteristic $\chi(O_\sigma)$ of the fixed point set:
\begin{equation}\label{LefTh}
\sum_i (-1)^i \left( b^i_+ - b^i_-\right) = \chi(O_\sigma) \qquad\mbox{where}\qquad b^i_\pm = \sum_{p+q=i} h^{p,q}_\pm \:.
\end{equation}
In our case $\chi(O_\sigma)=\chi(O7_1)+\chi(O7_2)+2\chi(O3)=75+3+2\cdot 1 = 80$. We know all the Betti numbers except $b^3_\pm$, i.e. $b^0_-=b^1_\pm=b^2_-=b^4_-=b^5_\pm=b_6^-=0$, $b^0_+=b^6_+=1$, $b^2_+=h^{1,1}=2$. The relation \eqref{LefTh} then gives $b^3_--b^3_+=74$, that in terms of $h^{1,2}_\pm$ becomes $h^{1,2}_--h^{1,2}_+=36$. Using also $h^{1,2}_++h^{1,2}_+=h^{1,2}=132$, we get $h^{1,2}_-=84$ and $h^{1,2}_+=48$.

\subsection{D7-brane setup}

An O7-plane has a D7-brane charge equal to $-8[D_{O7}]$. This must be canceled by the D7-branes. 
A D7-brane, wrapping the divisor $D$, can support a gauge invariant flux $\mathcal{F}\equiv F-\iota^\ast B$, where $F$ is the gauge flux and $\iota^\ast B$ is the pull-back of the NSNS two-form on $D$. The gauge flux must be quantised such that the Freed-Witten anomaly is canceled \cite{Freed:1999vc}, i.e. 
\begin{equation}
F+\frac{c_1(D)}{2}\in H^2(D,\mathbb{Z})
\end{equation}
(remember that in a CY $c_1(D)=-\iota^\ast D$, where $\iota^\ast$ is the pull-back map on $D$).

The $O7_2$ plane wraps a rigid divisor; hence its charge is canceled by four D7-branes (plus their orientifold images) wrapping $D_z$. If one takes 
\begin{equation}
B=\frac{D_z}{2}\:,
\end{equation} 
one can choose a quantised flux such that $\mathcal{F}=0$ on the D7-branes wrapping $D_z$.
This gives rise to an $SO(8)$ pure super Yang-Mills, supporting gaugino condensation (the zero flux prevents from breaking the gauge group to a subgroup with chiral spectrum). 


The D7-charge of the $O7_1$ plane can be cancelled by any D7-brane setup whose total homology class is $8[D_{O7_1}]$. We choose to work with a so called Whitney brane \cite{Collinucci:2008pf,Collinucci:2008sq}, an orientifold invariant D7-brane, wrapping the locus $\eta_{12,4}^2-w^2\chi_{18,6}=0$ in the homology class $8D_w$ (the subscript indicates the degree of the polynomial with respect to the $\mathbb{C}^\ast$ toric actions in \eqref{eq:model3dP8:weightm}). This brane can be seen as the result of the brane recombination between a D7-brane at $\eta_{12,4}-w\psi_{9,3}=0$ and its image D7-brane at $\eta_{12,4}+w\psi_{9,3}=0$. In this process the D-brane charges are conserved. This will be used in the next section to compute the D3-brane charge of the Whitney brane.

\subsection{D3-tadpole}

We now compute the contribution of the various objects to the D3-tadpole. 

We have two O3-planes, that give a total charge 
\begin{equation}
Q_{D3}^{O3}=2\times \left(- \frac{1}{2}\right)=-1 \:.
\end{equation}

The O7-planes contribute to the charge with $-\frac{\chi(O7)}{6}$, where $\chi(O7)$ is the Euler characteristic of the divisor wrapped by the O7-plane.
In our model
\begin{eqnarray}
Q_{D3}^{O7_1} &=& -\frac{\chi(D_w)}{6}= -\frac{75}{6}=-\frac{25}{2} \\
Q_{D3}^{O7_2} &=& -\frac{\chi(D_z)}{6}= -\frac{3}{6}=-\frac{1}{2} \\
\end{eqnarray}

Let us consider the D7-branes. The D3-charge of a single D7-brane wrapping a divisor $D$ and supporting the (gauge invariant) flux $\mathcal{F}$ is given by 
\begin{equation}
Q_{D3}^{D7} = - \frac{\chi(D)}{24} -\frac12\int_D	\mathcal{F}\wedge\mathcal{F}\:.
\end{equation}

The four D7-branes (plus their images) on top of the $O7_2$ divisor will then contribute as (remember that we chose $\mathcal{F}=0$)
\begin{equation}
Q_{D3}^{D7_2} = - (4+4)\times \frac{\chi(D)}{24} = -\frac{\chi(D_z)}{3}=-1 \:.
\end{equation}

We compute the D3-charge of the Whitney brane in the brane-image brane system. Both D7-branes wrap a divisor in the class $D_P=4D_w$. Since this class is even, the gauge flux must be integral. Moreover the pull-back of $B=\frac{D_z}{2}$ on $D_P$ vanishes (as $zw$ is in the SR-ideal).
The flux on one D7-brane is then $\mathcal{F}=n\,\iota^\ast D_u$ ($n\in\mathbb{Z}$), while on its image is $\mathcal{F}'=-n\,\iota^\ast D_u$. The D3-charge of the Whitney brane is equal to the sum of the D3-charges of the two D7-branes, i.e.
\begin{equation}
Q_{D3}^{D7_W} = -\frac{\chi(D_P)}{12} - \int_{X} n^2 D_P\wedge D_u\wedge D_u = -\frac{840}{12}-4n^2D_wD_u^2=-70 -4n^2\:.
\end{equation}
The range in which $n$ can vary is given by the condition that the Whitney brane is not forced to split into brane/image-brane\cite{Collinucci:2008pf,Collinucci:2008sq}, i.e. 
\begin{equation}
   \frac{[O7_1]}{2}-\frac{D_P}{2} \leq \mathcal{F} \leq   - \frac{[O7_1]}{2}+\frac{D_P}{2}
   \qquad\qquad\mbox{with}\qquad D_P-[O7_1]\geq 0\:.
\end{equation}
In our case this gives
\begin{equation}
-\frac{3}{2}D_w \leq nD_u \leq \frac{3}{2}D_w  \qquad\mbox{i.e.}\qquad  |n| \leq 4\:. 
\end{equation}
When we choose the biggest value $n=\pm 4$, we obtain 
\begin{equation}
Q_{D3}^{D7_W} = -70 -4\times 16 = -134\:.
\end{equation}

The total D3-charge of the O3/O7/D7 setup is 
\begin{equation}\label{ExplModelTotalD3charge}
Q_{D3}^{\rm O3/O7/D7}= -1 -\frac{25}{2} - \frac12 - 1 - 134 = -149\:.
\end{equation}

This is quite a big number. We will see that it is large enough to compensate the contribution from the  fluxes in the throat and it still leaves space to have bulk fluxes that stabilise  the bulk complex structure moduli.

\subsection{Warped throat with O3-planes at the tip}

In this section, we follow \cite{Garcia-Etxebarria:2015lif} to prove that there is a corner in the complex structure moduli space where a long throat is generated and moreover that at the tip of this throat there is a pair of O3-planes. We will eventually want to put an \ad3-brane on top of one of the two O3-planes, in order to give an explicit realisation of the construction in \cite{Kallosh:2015nia}.

We focus on the neighborhood of the O3-planes at $\xi=u_1=u_2=0$. 
If we plug these
relations inside the defining equation \eqref{CYeq2ndMd}, we get the equation
\begin{equation} \label{quadO3old}
w^4v -2b \,w^2z^2v^4 + c\, z^4 v^7 = 0
\end{equation}
in the one-dimensional ambient space with weight system
\begin{equation}
\begin{array}{c|ccc}
 & z &  v & w   \tabularnewline \hline 
    \mathbb{C}_1^* & 0  &  2  &  3  \tabularnewline 
    \mathbb{C}_2^* & 1  &  0  &  1 \tabularnewline
\end{array}
\end{equation}
We can first fix $z=1$ (if $z=0$ the equation above implies $w=0$ as well, but they cannot vanish simultaneously), going to the one dimensional ambient space
\begin{equation}
\begin{array}{c|cc}
 &   v & w   \tabularnewline \hline 
    \mathbb{C}^* &   2  &  3  \tabularnewline 
\end{array}
\end{equation}
Then, we note that $v\neq 0$, as $v$ cannot vanish together with $u_1$ and $u_2$ (see the SR-ideal). 
Fixing $v=1$, we are left with the ambient space $\mathbb{C}/\mathbb{Z}_2$ ($\mathbb{Z}_2$: $w\mapsto -w$) and the equation
\begin{equation}\label{quadO3}
w^4 -2b \, w^2 +c =0 \:.
\end{equation}
We then explicitly see why we obtain two O3's (the equation is solved by $w^2=\gamma_i$, with $\gamma_i$ the zeroes of the quadratic equation, the solutions $w=\pm\sqrt{\gamma_i}$ are identified by the $\mathbb{Z}_2$ orbifold action).
These two points come on top of each other when the discriminant of the quadratic equation is zero, i.e. when 
\begin{equation}
  b^2-c =0 \:.
\end{equation}
So, let us redefine 
\begin{equation}
c \equiv b^2 + \delta \:.
\end{equation}
When $\delta=0$ the two O3-planes go on top of each other at the point $w^2-b z^2 =0$. When $\delta$ is small they are very close to each other. 

As a next step we need to check that in the limit $\delta\rightarrow 0$ a conifold singularity is generated on the CY three-fold at the point $\xi=u_1=u_2= w^2-b z^2=0$. 
It is enough to consider a small neighborhood of the points $\xi=u_1=u_2= (w^2-b z^2)^2+\delta=0$. Here  $v$ and $z$ still do not vanish and we can fix them to $1$.
In this local patch, the equation defining the CY three-fold becomes (in an ambient space $\mathbb{C}^4/\mathbb{Z}_2$)
\begin{equation}\label{conifeqaij}
 -\xi^2 + \sum_{i,j=1,2} a_{ij}u_iu_j +    (w^2-b)^2  + ... = \delta 
\end{equation}
where `$...$' are monomials that vanish faster that the quadratic ones at $\xi=u_1=u_2=w^2-b=0$ (we keep $\delta$ small). The equation \eqref{conifeqaij} 
describes a deformed conifold singularity. This becomes manifest if we diagonalise the quadratic form $a_{ij}$ and write the equation  as\footnote{It seems that when $\delta\rightarrow 0$ we get two conifold singularities, one at $w=\sqrt{b}$ and one at $w=-\sqrt{b}$. However, these points are identified by the orbifold action. Note also that the conifold point is far apart from the orbifold singularity of the ambient space, that anyway never belongs to the CY three-fold.}
\begin{equation}
 -\xi^2 + u_1^2 + u_2^2 +    (w^2-b)^2  + ... = \delta  \:.
\end{equation}
We have then shown what we aimed, i.e. that when taking $\delta\rightarrow 0$ we obtain a conifold singularity. $\delta$ is the complex structure modulus that controls the size of the $S^3$ at the tip of the throat.

We need finally to find how the involution is acting on the conifold. By carefully inspecting the gauge fixing $v=z=1$, one discovers that in the local patch we are considering the involution becomes 
\begin{equation}
 \xi \mapsto - \xi \,,\qquad u_1 \mapsto - u_1 \,,\qquad u_2 \mapsto - u_2 \,,
\end{equation}
that perfectly reproduces the geometric action required for the
retrofitting of a nilpotent Goldstino sector \cite{Garcia-Etxebarria:2015lif}.

\medskip

Let us finish with a comment on the D3-charge. Remember
first that we have two O3-planes at the tip of the throat. The D3-charge of the system of the
O7-planes and tadpole cancelling D7-branes is integral. As discussed in \cite{Garcia-Etxebarria:2015lif}, one can choose for
instance to put a stuck D3 at one of the O3$^-$ points on the
contracting $S^3$, and a stuck $\overline{D3}$ on the other O3$^-$ on this same
$S^3$. This pair of stuck branes does not contribute to the D3-charge \eqref{ExplModelTotalD3charge}. As the D-branes are stuck at the O3-planes, there is no perturbative decay channel between the D3 and the \ad3 (if the c.s. are fixed such that the $S^3$ has finite size) \cite{Garcia-Etxebarria:2015lif}.

\subsection{Moduli stabilisation}

We have studied moduli stabilisation in Section~\ref{sec:2steps}. We use those results in this section. We will take the parameter $\xi$ in the K\"ahler potential \eqref{eq:Kpotential} as given by $ \xi \equiv-\frac{\chi(X) \zeta(3)}{4(2 \pi)^{3}}$, without including the correction of \cite{Minasian:2015bxa}; the results we find do not differ sensibly from those obtained by adding that correction (the Euler characteristic of this model is $\chi(X)=-260$; with the correction of \cite{Minasian:2015bxa} the Euler characteristic would be  effectively shifted to $260-36=224$). From equation \eqref{eq:volExplModel} we read $\kappa_s=\frac{\sqrt{2}}{9}$. Moreover the non-perturbative superpotential is generated by gaugino condensation on the $SO(8)$ D7-brane stack wrapping the divisor $D_z$, i.e. $a=\frac{\pi}{3}$.

Applying the methods of the numerical analysis of Section~\ref{Sec:conditions1-3}, one finds that the only allowed values of the throat flux D3-charge are $MK=\{88, 92, 125\}$.  
We choose the values of the  parameters $W_0,g_s,M,K$ that produce a model with $MK=88$, i.e.
\begin{equation} 
	W_0=23; \qquad g_s=0.23; \qquad M=22; \qquad K=4 \:.
\end{equation}
Notice that $g_s M=5.02$, in agreement with the requirement (3) of Section~\ref{Sec:conditions1-3}.
	
In this case the moduli are stabilised at:
\[\tau_s=7.6;\ \tau_b=704;\ \zeta=0.005\]
which means, in particular that the volume  $\vol\simeq\tau_b^{3/2}=18698\gg 1$ (in Einstein frame; in string frame this corresponds to $\vol_s=2036$) as needed.
These values correspond to a de~Sitter minimum with $V_{min}\simeq 10^{-12}$.   

\begin{figure}[t!]
	\includegraphics[width=.6\textwidth]{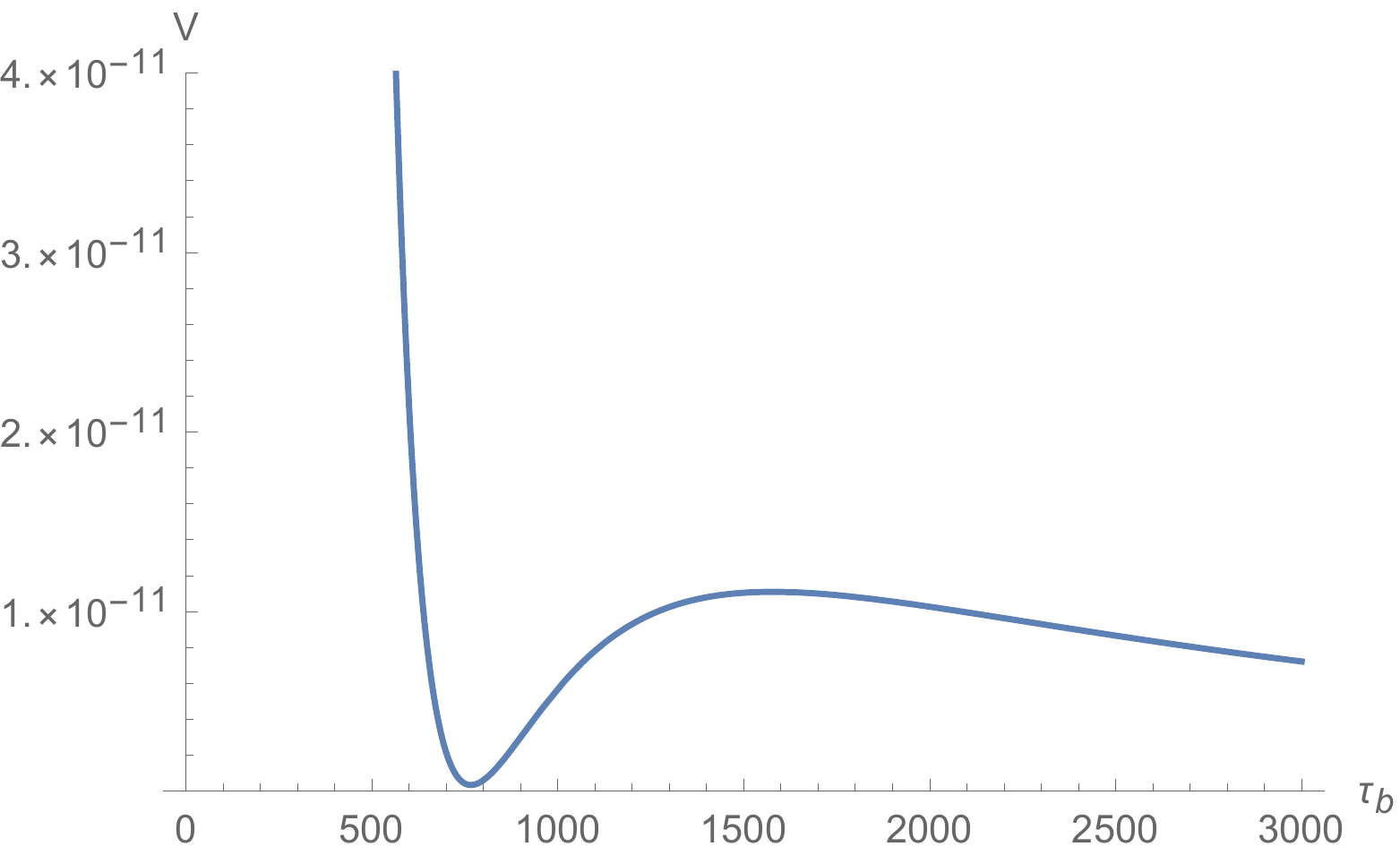}
	\caption{\label{fig:weakestDirectionNEW260} The total scalar potential along its weakest direction $(\tau_s=\tau_s^{min})$.}
\end{figure}

The value of the scalar potential at the minimum can be decreased (without breaking the other constraints) by fine-tuning $W_0$: the finer the tuning, the smaller the potential at the minimum. 
The masses of the moduli are:
	\[m_1^2=3.7\times 10^{-5}M_p,\qquad m_{3}^2=9.5\times10^{-13}M_p,\qquad m_2^2=8.7\times 10^{-10}M_p,\]
showing the predicted hierarchy ($m_1,m_3$ are approximately the masses of the moduli $\tau_s$ and $\tau_b$ at the LVS minimum).

The other scales are:
	\[M_s=0.009M_p\simeq 3.6  M_{KK}^{bulk}, \,\,\,\, M_{KK}^{bulk}=0.002 M_p\simeq 6 m_{3/2},\,\,\,\,
	m_{3/2}= 4.3\times 10^{-4}M_p,\]
hence they respect the correct hierarchy.
As predicted in Section~\ref{Sec:Constraints}, all the remaining consistency conditions are fulfilled.

\begin{figure}[t]
	\includegraphics[width=.6\textwidth]{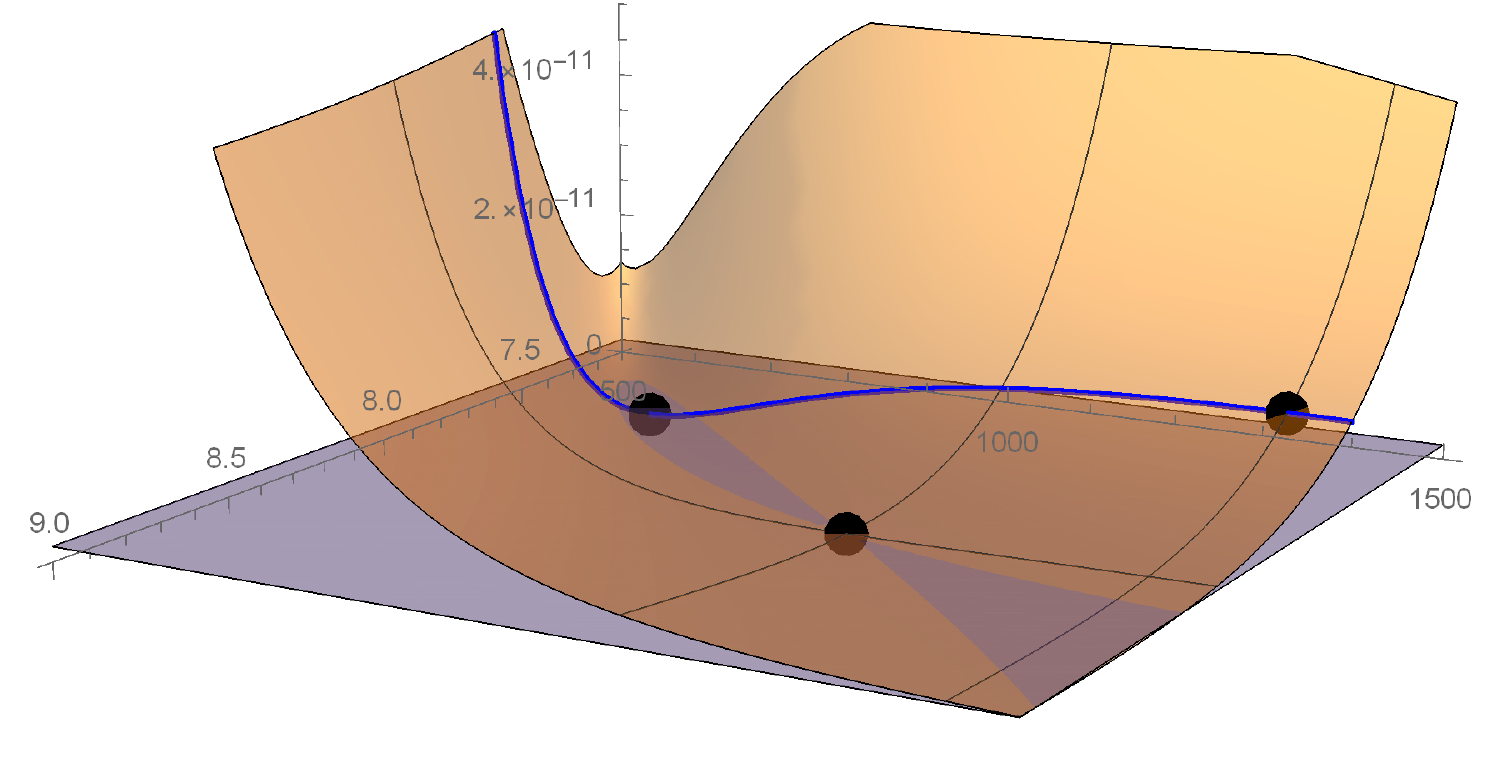}
	\caption{\label{fig:ScalarPotentialNEW260} Total scalar potential  as a function of $\tau_s$ and $\tau_b$. The black dots correspond to the de Sitter minimum, the saddle point and the maximum along the weakest direction (the blue curve). The grey plan at $V=V_{saddle}$ highlights the unstable direction for the saddle point.}
\end{figure}

To have a qualitative idea of the potential, we fix all the moduli but $\tau_s,\tau_b$ at their vacuum value. The resulting two-moduli potential,\footnote{The 3-moduli analysis, including $\zeta$, gives analogous results.} 
beside the minimum, has a second stationary point at  
	\[(\tau_s,\tau_b)=(8.2, 1123)\]
that turns out to be a saddle point.\footnote{The global maximum of $V$ is, instead on the boundary of the \K cone $(\tau_b=\tau_s=\tau_s^{min})$ where $V_{max}=0.06$.}
If we furthermore keep $\tau_s$ at its vev, the resulting one modulus potential is plot in  Figure~\ref{fig:weakestDirectionNEW260}: it has a maximum at  $\tau_b=1433\equiv\tau_b^{max}$, with $V_{\tau_b^{max}}=1.2\times 10^{-11}$ \footnote{This gives an estimate of the barrier for tunneling from the minimum to decompactification. As discussed in \cite{deAlwis:2013gka} the decay rate in LVS can then be estimated to be of order $\Gamma\sim e^{-\vo^3}\sim e^{-10^{36}}$ in Planck units which, similar to the KKLT case, implies a very stable vacuum with a lifetime smaller than the recurrence time.}.
The 2-moduli scalar potential is plotted in  Fig. \ref{fig:ScalarPotentialNEW260}, where all the significant points and the 1-modulus direction  are highlighted.

\section{Conclusions}\label{sec:Concl}

In this paper we have studied the moduli scalar potential for a type IIB compactification in  LVS regime, with a dS minimum realised by the introduction of an \ad3-brane at the tip of a warped throat. We assumed that all complex structure moduli (except the one relevant  for the throat) and the dilaton were fixed by fluxes as in GKP and we concentrated on the scalar potential for the throat complex structure modulus and the K\"ahler moduli.

We have seen that de Sitter minima exist only for a limited range of values of the warp factor controlling the uplift term in the potential: increasing the uplift term from the value realising a Minkowski minimum, the potential soon develops unstable directions. For these values of the warp factor the minimum is mildly shifted with respect to the LVS minimum.

Since the warp factor depends on the 3-form fluxes $M,K$ along the throat, the limited range of the warp factor constrains strongly the ratio of the integers $M$ and $K$. If one moreover requires the validity of the KS approximation, i.e. $g_s|M| \gg 1$, one has a lower bound on $M$ that, due to the fixed ratio, puts also a lower bound on $K$ and in particular on the product $MK$. This means that the D3-charge generated from the fluxes in the throat is bounded from below. Unfortunately this bound is typically large. 

In type IIB compactifications the flux contribution to the D3-charge, coming form $MK$ plus the contribution of the bulk fluxes necessary for complex structure moduli stabilisation, must be cancelled by localised sources. D7-branes and O3/O7-planes give in fact a negative contribution to the D3-charge. However,  having a large negative D3-charge, necessary to compensate the lower bound on $MK$, is challenging even if it is  possible.

In this paper we estimate the bound on $MK$ as one varies the parameters of the effective field theory. 
We conclude that this bound is minimal exactly at the boundary where the approximation used to obtain the minimum can be trusted.

In the last part of the article, we construct an explicit compact model in perturbative type IIB string theory, where an explicit throat supporting O3-planes is realised. This allows to describe the \ad3-brane degrees of freedom by the introduction of a single nilpotent chiral superfield.
We have shown in this model how one can generate a relatively large negative D3-charge by a proper D7-brane background, also in perturbative type IIB string theory. However, to satisfy the D3-tadpole cancellation condition we have to take a not-so-large value of $g_sM$, i.e. $g_sM\sim 5$ that is bigger than $1$ but it is not so big. Moreover also the string coupling should be taken small but not-so-small, i.e. $g_s\sim 0.23$. One may decrease the string coupling (improving  string perturbation theory), but this would lead to decreasing $g_sM$, worsening the KS supergravity approximation. 

In summary we conclude that within a class of models which are simple enough to be explicit but rich enough to include all the ingredients of  moduli stabilisation, we were able to provide a concrete de Sitter model from antibrane uplift satisfying all consistency constraints while also having small expansion parameters. We see this as a small step towards a more systematic realisation of de Sitter vacua in string compactifications. As expected, this was not a simple task  but having been able to find a vacuum in a regime in which the expansion parameters are small is encouraging. We are confident that with more elaborate compactifications, including chiral matter,  de Sitter solutions with  small expansion parameters, are achievable. We let a systematic search for the future.


\section*{Acknowledgements}

We would like to thank Michele Cicoli, Luca Martucci, Andrea Sangiovanni and Andreas Schachner for enlightening discussions.
The work of R.V.~is partially supported by ``Fondo per la Ricerca di Ateneo - FRA 2018'' (UniTS) and by INFN Iniziativa Specifica ST\&FI. The work of C.C.~is partially supported by INFN Iniziativa Specifica ST\&FI. The work of FQ has been partially supported by STFC consolidated grants ST/P000681/1, ST/T000694/1.

\newpage

\providecommand{\href}[2]{#2}\begingroup\raggedright\endgroup

\end{document}